\documentclass[lettersize,journal]{IEEEtran}
\usepackage[
    pdfborder={0 0 0},
    colorlinks=false
]{hyperref}
\IEEEoverridecommandlockouts
\pdfoutput=1
\usepackage[pdftex]{}
\usepackage[acronym]{glossaries}
\usepackage[nohyperlinks,nolist]{acronym}
\usepackage{cite}
\usepackage{amsmath,amssymb,amsfonts}
\usepackage{algorithmic}
\usepackage{graphicx}
\usepackage{textcomp}
\usepackage{xcolor}
\usepackage{orcidlink}
\usepackage{amssymb}
\usepackage{soul}
\usepackage{cancel}
\usepackage{tcolorbox}     
\usepackage{cuted}
\usepackage{subcaption}
\usepackage{float}
\usepackage{amsthm}
\usepackage[bottom]{footmisc}
\usepackage{atbegshi}
\usepackage{picture}
\newcommand{\licensenotice}{
    \AtBeginShipoutNext{
        \AtBeginShipoutUpperLeft{
            \put(\dimexpr\paperwidth/2\relax, -\dimexpr\paperheight-0.65cm\relax){ 
                \makebox[0pt][c]{
                    \small This work is licensed under a Creative Commons Attribution 4.0 International License. For more information, see \href{https://creativecommons.org}{creativecommons.org}.
                }
            }
        }
    }
}

\renewcommand{\footnoterule}{%
  \kern -3pt
  \hrule width \columnwidth height 0.4pt
  \kern 2.6pt
}
\newtheoremstyle{assumpstyle}      
  {\topsep}
  {\topsep}
  {\normalfont}
  {}
  {\bfseries\itshape}
  {.}
  { }
  {\thmname{#1}\thmnumber{ #2}\thmnote{ (#3)}}

\theoremstyle{assumpstyle}

\def\BibTeX{{\rm B\kern-.05em{\sc i\kern-.025em b}\kern-.08em
    T\kern-.1667em\lower.7ex\hbox{E}\kern-.125emX}}
\usepackage{enumitem}
\usepackage[dvipsnames]{xcolor}
            
\begin{document}


    
    
    
\newacronym{3GPP}{3GPP}{3rd Generation Partnership Project}

\newacronym{5G}{5G}{5th Generation}
\newacronym{5G-ACIA}{5G-ACIA}{5G Alliance for Connected Industries and Automation}
\newacronym{5GC}{5GC}{5G Core}
\newacronym{5QI}{5QI}{5G QoS Identifier}

\newacronym{6G}{6G}{6th Generation}

\newacronym{AF}{AF}{Application Function}
\newacronym{AI}{AI}{Artificial Intelligence}
\newacronym{AMBR}{AMBR}{Aggregate Maximum Bit Rate}
\newacronym{AMR}{AMR}{Autonomous Mobile Robot}
\newacronym{ARP}{ARP}{Allocation and Retention Priority}
\newacronym{ATS}{ATS}{Asynchronous Time-aware Shaper}
\newacronym{AVB}{AVB}{Audio Video Bridging}

\newacronym{BC}{BC}{Boundary Clock}
\newacronym{BE}{BE}{Best-Effort}
\newacronym{BLER}{BLER}{Block Error Rate}
\newacronym{BS}{BS}{Base Station}

\newacronym{C2C}{C2C}{Controller-to-controller}
\newacronym{C2D}{C2D}{Controller-to-device}
\newacronym{CCDF}{CCDF}{Complementary Cumulative Distribution Function}
\newacronym{CDF}{CDF}{Cumulative Distribution Function}
\newacronym{CF}{CF}{Correction Field}
\newacronym{CG}{CG}{Configured Grant}
\newacronym{CNC}{CNC}{Centralized Network Configuration}
\newacronym{CQI}{CQI}{Channel Quality Indicator}
\newacronym{CPE}{CPE}{Customer Premise Equipment}
\newacronym{CPS}{CPS}{Cyber-Physical System}
\newacronym{CRAS}{CRAS}{Connected Robotics and Autonomous Systems}
\newacronym{CSMA}{CSMA}{Carrier Sense Multiple Access}

\newacronym{D2Cmp}{D2Cmp}{Device-to-computer}
\newacronym{DCI}{DCI}{Downlink Control Information}
\newacronym{DC}{DC}{Delay-Critical}
\newacronym{DG}{DG}{Dynamic Grant}
\newacronym{DL}{DL}{Downlink}
\newacronym{DN}{DN}{Data Network}
\newacronym{DNN}{DNN}{Data Network Name}
\newacronym{DRB}{DRB}{Data Radio Bearer}
\newacronym{DRL}{DRL}{Deep Reinforcement Learning}
\newacronym{DSCP}{DSCP}{Differentiated Services Code Point}
\newacronym{DS-TT}{DS-TT}{Device-side Translator}
\newacronym{D-TDD}{D-TDD}{Dynamic TDD}

\newacronym{E2E}{E2E}{End-to-End}
\newacronym{eMBB}{eMBB}{enhanced Mobile BroadBand}

\newacronym{FD}{FD}{Field Device}
\newacronym{FDD}{FDD}{Frequency Division Duplex}
\newacronym{FIFO}{FIFO}{First-In First-Out}
\newacronym{FPGA}{FPGA}{Field-Programmable Gate Array}
\newacronym{FRER}{FRER}{Frame Replication and Elimination for Reliability}

\newacronym{GCD}{GCD}{Greatest Common Divisor}
\newacronym{GCL}{GCL}{Gate Control List}
\newacronym{GM}{GM}{Grand Master}
\newacronym{gNB}{gNB}{next generation Node B}
\newacronym{GNSS}{GNSS}{Global Navigation Satellite System}
\newacronym{gPTP}{gPTP}{generalized Precision Time Protocol}
\newacronym{GTP}{GTP}{GPRS Tunneling Protocol}

\newacronym{HIL}{HIL}{Hardware-in-the-Loop}
\newacronym{HARQ}{HARQ}{Hybrid Automatic Repeat Request}
\newacronym{HP}{HP}{High Priority}
\newacronym{HMP}{HMP}{High-Medium Priority}

\newacronym{ICI}{ICI}{Inter-Cycle Interference}
\newacronym{IFI}{IFI}{Inter-Flow Interference}
\newacronym{IEEE}{IEEE}{Institute of Electrical and Electronics Engineers}
\newacronym{IIoT}{IIoT}{Industrial Internet of Things}
\newacronym{IoT}{IoT}{Internet of Things}

\newacronym{L2C}{L2C}{Line controller-to-controller}
\newacronym{LCM}{LCM}{Least Common Multiple}
\newacronym{LMP}{LMP}{Low-Medium Priority}
\newacronym{LP}{LP}{Low Priority}
\newacronym{LTE}{LTE}{Long Term Evolution}

\newacronym{MAC}{MAC}{Medium Access Control}
\newacronym{MBS}{MBS}{Multicast–Broadcast Services}
\newacronym{MBSFN}{MBSFN}{Multicast-Broadcast Single-Frequency Network}
\newacronym{MB-UPF}{MB-UPF}{Multicast/Broadcast UPF}
\newacronym{MCS}{MCS}{Modulation and Coding Scheme}
\newacronym{MES}{MES}{Manufacturing Execution System}
\newacronym{MNO}{MNO}{Mobile Network Operator}
\newacronym{MTU}{MTU}{Maximum Transmission Unit}
\newacronym{Multi-TRP}{Multi-TRP}{Multiple Transmission and Reception Point}
\newacronym{MS}{MS}{Master}

\newacronym{NIC}{NIC}{Network Interface Card}
\newacronym{NW-TT}{NW-TT}{Network-side Translator}
\newacronym{NR}{NR}{New Radio}
\newacronym{NTP}{NTP}{Network Time Protocol}
\newacronym{NPN}{NPN}{Non-Public Network}

\newacronym{OFDMA}{OFDMA}{Orthogonal Frequency-Division Multiple Access}
\newacronym{OFDM}{OFDM}{Orthogonal Frequency-Division Multiplexing}
\newacronym{O-RAN}{O-RAN}{Open Radio Access Network}


\newacronym{P2P}{P2P}{Peer-to-Peer}
\newacronym{PCP}{PCP}{Priority Code Point}
\newacronym{PDCCH}{PDCCH}{Physical Downlink Control Channel}
\newacronym{PDCP}{PDCP}{Packet Data Convergence Protocol}
\newacronym{PDF}{PDF}{Probability Density Function}
\newacronym{PDR}{PDR}{Packet Detection Rule}
\newacronym{PDSCH}{PDSCH}{Physical Downlink Shared Channel}
\newacronym{PDU}{PDU}{Packet Data Unit}
\newacronym{PLC}{PLC}{Programmable Logic Controller}
\newacronym{PPS}{PPS}{Pulse Per Second}
\newacronym{PRACH}{PRACH}{Physical Random Access Channel}
\newacronym{PTP}{PTP}{Precision Time Protocol}
\newacronym{PUCCH}{PUCCH}{Physical Uplink Control Channel}
\newacronym{PUSCH}{PUSCH}{Physical Uplink Shared Channel}

\newacronym{QAM}{QAM}{Quadrature Amplitude Modulation}
\newacronym{QoE}{QoE}{Quality of Experience}
\newacronym{QoS}{QoS}{Quality of Service}
\newacronym{QFI}{QFI}{QoS Flow ID}

\newacronym{RAN}{RAN}{Radio Access Network}
\newacronym{RB}{RB}{Resource Block}
\newacronym{RLC}{RLC}{Radio Link Control}
\newacronym{RRC}{RRC}{Radio Resource Control}
\newacronym{RTC1}{RTC1}{Real Time Class 1}

\newacronym{SCADA}{SCADA}{Supervisory Control and Data Acquisition}
\newacronym{SCS}{SCS}{Sub-Carrier Spacing}
\newacronym{SDR}{SDR}{Software Defined Radio}
\newacronym{SFP}{SFP}{Small Form-factor Pluggable}
\newacronym{SMF}{SMF}{Session Management Function}

\newacronym{SINR}{SINR}{Signal-to-Interference-plus-Noise Ratio}
\newacronym{SPS}{SPS}{Semi-Persistent Scheduling}
\newacronym{SST}{SST}{Slice/Service Type}
\newacronym{S-NSSAI}{S-NSSAI}{Single Network Slice Selection Assistance Information}
\newacronym{SL}{SL}{Slave}

\newacronym{TAS}{TAS}{Time-Aware Shaper}
\newacronym{TC}{TC}{Transparent Clock}
\newacronym{TDD}{TDD}{Time Division Duplex}
\newacronym{TDMA}{TDMA}{Time-Division Multiple Access}
\newacronym{TSC}{TSC}{Time-Sensitive Communication}
\newacronym{TSN}{TSN}{Time-Sensitive Networking}
\newacronym{TTI}{TTI}{Transmission Time Interval}

\newacronym{UDP}{UDP}{User Datagram Protocol}
\newacronym{UE}{UE}{User Equipment}
\newacronym{UFTP}{UFTP}{UDP-based File Transfer Protocol}
\newacronym{UL}{UL}{Uplink}
\newacronym{UPF}{UPF}{User Plane Function}
\newacronym{URLLC}{URLLC}{Ultra-Reliable and Low-Latency Communications}
\newacronym{uRLLC}{uRLLC}{ultra-Reliable and Low-Latency Communications}

\newacronym{VLAN}{VLAN}{Virtual Local Area Network}
\newacronym{VNF}{VNF}{Virtualized Network Function}
\newacronym{VNI}{VNI}{Virtual Network Identifier}
\newacronym{vPLC}{vPLC}{virtualized PLC}
\newacronym{VTEP}{VTEP}{VxLAN Tunnel End Point}
\newacronym{VxLAN}{VxLAN}{Virtual Extensible LAN}
\newacronym{vBBU}{vBBU}{virtual Baseband Unit}

\newacronym{ZWSL}{ZWSL}{Zero-Wait-at-SL}

\title{Impact of 5G Latency and Jitter on TAS Scheduling in a 5G-TSN Network: An Empirical Study}

\author{
\IEEEauthorblockN{Pablo Rodriguez-Martin\orcidlink{0000-0002-7345-8923}, Oscar Adamuz-Hinojosa\orcidlink{0000-0002-7797-1150}, Pablo Muñoz\orcidlink{0000-0002-3265-5728}, Julia Caleya-Sanchez\orcidlink{0000-0002-1703-8641}, 
Pablo Ameigeiras\orcidlink{0000-0002-4572-3902}}

\thanks{This work has been financially supported by the Ministry for Digital Transformation and the Civil Service of the Spanish Government through TSI-063000-2021-28 (6G-CHRONOS) project, and by the European Union through the Recovery, Transformation and Resilience Plan - NextGenerationEU. Additionally, this publication is part of grant PID2022-137329OB-C43 funded by MICIU/AEI/10.13039/501100011033 and ERDF/EU, and part of FPU Grant 21/04225 funded by the Spanish Ministry of Universities. Funding for open access charge: Universidad de Granada / CBUA.

The authors are with the Department of Signal Theory, Telematics and Communications; and with the Research Center on Information and Communication Technologies, both from the University of Granada, Granada, Spain. E-mails: \texttt{\{pablorodrimar, oadamuz, pabloml, jcaleyas, pameigeiras\}@ugr.es}.}
}

\licensenotice
\maketitle
\setcounter{page}{1}

\begin{abstract}
Deterministic communications are essential to meet the stringent delay and jitter requirements of Industrial Internet of Things (IIoT) services. IIoT increasingly demands wide-area wireless mobility to support Autonomous Mobile Robots (AMR) and dynamic workflows. Integrating Time-Sensitive Networking (TSN) with 5G private networks is emerging as a promising approach to fulfill these requirements. In this architecture, 5G provides wireless access for industrial devices, which connect to a TSN backbone that interfaces with the enterprise edge/cloud, where IIoT control and computing systems reside. TSN achieves bounded latency and low jitter using IEEE 802.1Qbv Time-Aware Shaper (TAS), which schedules the network traffic in precise time slots. However, the stochastic delay and jitter inherent in 5G disrupt TSN scheduling, requiring careful tuning of TAS parameters to maintain end-to-end determinism. This paper presents an empirical study evaluating the impact of 5G downlink delay and jitter on TAS scheduling using a testbed with TSN switches and a commercial 5G network. Results show that guaranteeing bounded latency and jitter requires careful setting of TAS transmission window offset between TSN switches based on the measured 5G delay bounded by a high order p-th percentile. Otherwise, excessive offset may cause additional delay or even a complete loss of determinism. 
\end{abstract}


\section{Introduction}
\gls{IIoT} enables tightly integrated \glspl{CPS}, which are critical for manufacturing automation in modern Industry 4.0. These systems demand deterministic, low-latency communication to guarantee safe and predictable operation in dynamic industrial environments~\cite{Groshev2021}. Among the most demanding \gls{IIoT} applications are \gls{CRAS}, including \glspl{AMR}, drones, and intelligent agents. These systems rely on precise coordination between sensing, computing, and actuation, and are highly sensitive to communication delays and jitter~\cite{Saad20}.

To meet these demands, \gls{TSN} standards define mechanisms that enable deterministic communication over wired Ethernet infrastructures~\cite{ieee8021q}. One of the key components of \gls{TSN} is the IEEE 802.1Qbv \gls{TAS}, which operates at the output ports of \gls{TSN} switches. \gls{TAS} enforces scheduled access to the transmission medium by periodically opening and closing gates that control the egress of packets from different traffic queues. By precisely determining when each queue is allowed to transmit, \gls{TAS} ensures bounded delay and low jitter for selected traffic classes. This deterministic behavior is essential to support time-critical \gls{IIoT} applications that require guaranteed communication performance~\cite{ieee8021qbv}. Nevertheless, \gls{TSN}'s reliance on wired infrastructure limits mobility and flexibility, especially in complex industrial settings.

To overcome these limitations, \gls{5G} mobile networks offer mobility, flexibility, wide-area coverage, and \gls{uRLLC} capabilities, which have sparked significant interest in integrating \gls{5G} with \gls{TSN} for industrial scenarios~\cite{5GACIA-whitepaperI}. In this paradigm, industrial end devices such as robots and production line equipment connect wirelessly to the network via the \gls{5G} system. The \gls{5G} network provides access to a wired \gls{TSN} backbone composed of \gls{TSN} switches connected to edge computing platforms hosting \gls{IIoT} control functions. This integration aims to combine \gls{5G} mobility and coverage with \gls{TSN} determinism. However, the stochastic nature of \gls{5G}, characterized by variable delay in the radio and core segments, disrupts the strict timing required by \gls{TAS}. This variability challenges the achievement of \gls{E2E} deterministic communication.

To address these challenges, \gls{TAS} configurations must be carefully adapted to maintain synchronized transmissions across \gls{TSN} switches. In particular, these configurations must compensate for the delay variability introduced by the \gls{5G} system while avoiding excessive buffering, added latency, or bandwidth inefficiencies. Ensuring proper alignment of transmission windows is essential to preserve the deterministic guarantees required by time-sensitive \gls{IIoT} applications.

\textbf{Literature Review.} The \gls{5G}-\gls{TSN} integration has drawn substantial research interest. Prior works have explored architectures where \gls{5G} functions as a logical \gls{TSN} switch and have proposed solutions for time synchronization and \gls{QoS} mapping between domains. Simulation studies have also evaluated \gls{TAS} scheduling and jitter mitigation; however, these typically rely on idealized wireless models. Although such studies have advanced the understanding of \gls{5G}-\gls{TSN} integration, critical challenges remain in tuning \gls{TAS} parameters to compensate for realistic \gls{5G} delay and jitter dynamics. In particular, there is a lack of experimental validation under commercial \gls{5G} conditions. For interested readers, a detailed literature review is provided in Section~\ref{sec:related_works}. 

\textbf{Contributions.} This article analyzes the impact of \gls{5G}-induced delay and jitter on the operation of the IEEE 802.1Qbv \gls{TAS} in an integrated \gls{5G}-\gls{TSN} network, focusing on the configuration of \gls{TAS} scheduling parameters to accommodate a delay-critical traffic flow. The main contributions are:

\begin{itemize}
    \item[C1] We provide a detailed analysis of the delay components involved in the transmission of packets between adjacent \gls{TAS}-enabled \gls{TSN} switches interconnected via a \gls{5G} network. This analysis characterizes how \gls{5G}-induced delays and jitter interact with \gls{TAS} parameters, and quantifies their impact on \gls{E2E} latency performance.
    \item[C2] Based on this analysis, we identify the conditions under which deterministic communication can be achieved in \gls{5G}-\gls{TSN} networks. We thoroughly investigate the resulting scenarios arising from different \gls{TAS} parameter configurations and provide general configuration guidelines to ensure deterministic behavior.
    \item[C3] We implement an experimental testbed integrating a commercial private \gls{5G} network and \gls{TAS}-enabled \gls{TSN} switches, enabling real-world evaluation of \gls{TAS} configurations under realistic conditions. The testbed is used to assess the impact of \gls{5G} delay and jitter on specific \gls{TAS} settings under representative network scenarios.
\end{itemize}

This article builds upon our previous conference work~\cite{rodriguez2025}, which presented an initial testbed-based study of \gls{TAS} scheduling in integrated \gls{5G}-\gls{TSN} environments. In this extended version, we provide a more comprehensive theoretical and experimental analysis of the impact of \gls{5G}-induced delay and jitter on \gls{TAS} operation. We identify and characterize critical scenarios arising from different \gls{TAS} parameter configurations. Furthermore, we derive general configuration guidelines and formally establish the conditions required to guarantee deterministic \gls{E2E} performance in \gls{5G}-\gls{TSN} networks.

Our results show that guaranteeing bounded latency and jitter requires configuring the \gls{TAS} transmission window offset between \gls{TSN} switches based on the maximum observed \gls{5G} delay, estimated using a high-percentile delay metric. While increasing this offset helps to absorb delay variability, it also increases \gls{E2E} latency. Moreover, if the offset becomes excessively large, it may cause misalignment between the transmission windows of \gls{TSN} switches, thereby violating the deterministic behavior. Additionally, to ensure that packets always arrive within their assigned transmission windows, the \gls{TAS} cycle period should be greater than the sum of the peak-to-peak jitter introduced by \gls{5G} and the transmission window duration. Finally, we see how additional traffic flows with the same priority may also increase \gls{5G} delay and jitter. Similarly, if the \gls{5G} network lacks proper isolation between traffic types, flows with lower priority can contribute to latency and jitter degradation. Such cases require recalculating \gls{TAS} parameters.

\textbf{Paper Outline.} The paper is organized as follows: Section~\ref{sec:Background} covers background on industrial \gls{5G}-\gls{TSN} networks and \gls{TAS}. Section~\ref{sec:SystemModel} presents the system model. Section~\ref{sec:jitter} analyzes \gls{5G} delay and jitter impact on \gls{TAS}. Section~\ref{sec:Testbed} describes the testbed and the experimental setup. Section~\ref{sec:Results} reports performance results. Section~\ref{sec:related_works} reviews related work. Section~\ref{sec:Conclusions} outlines the key conclusions and future work.

\section{Background on Industrial \gls{5G}-\gls{TSN} Networks}\label{sec:Background}
This section overviews \gls{5G}-\gls{TSN} networks in Industry 4.0. First, we introduce the main network segments and key characteristics of industrial applications. Then, we discuss \gls{QoS} traffic management and the \gls{TAS} mechanism. Finally, we highlight time synchronization for deterministic communications.

\subsection{\gls{5G}-\gls{TSN} Network Segments for Industry Automation}\label{sec:network_segments}

\begin{figure}[b!]
    \centering
   \includegraphics[width=0.98\columnwidth]{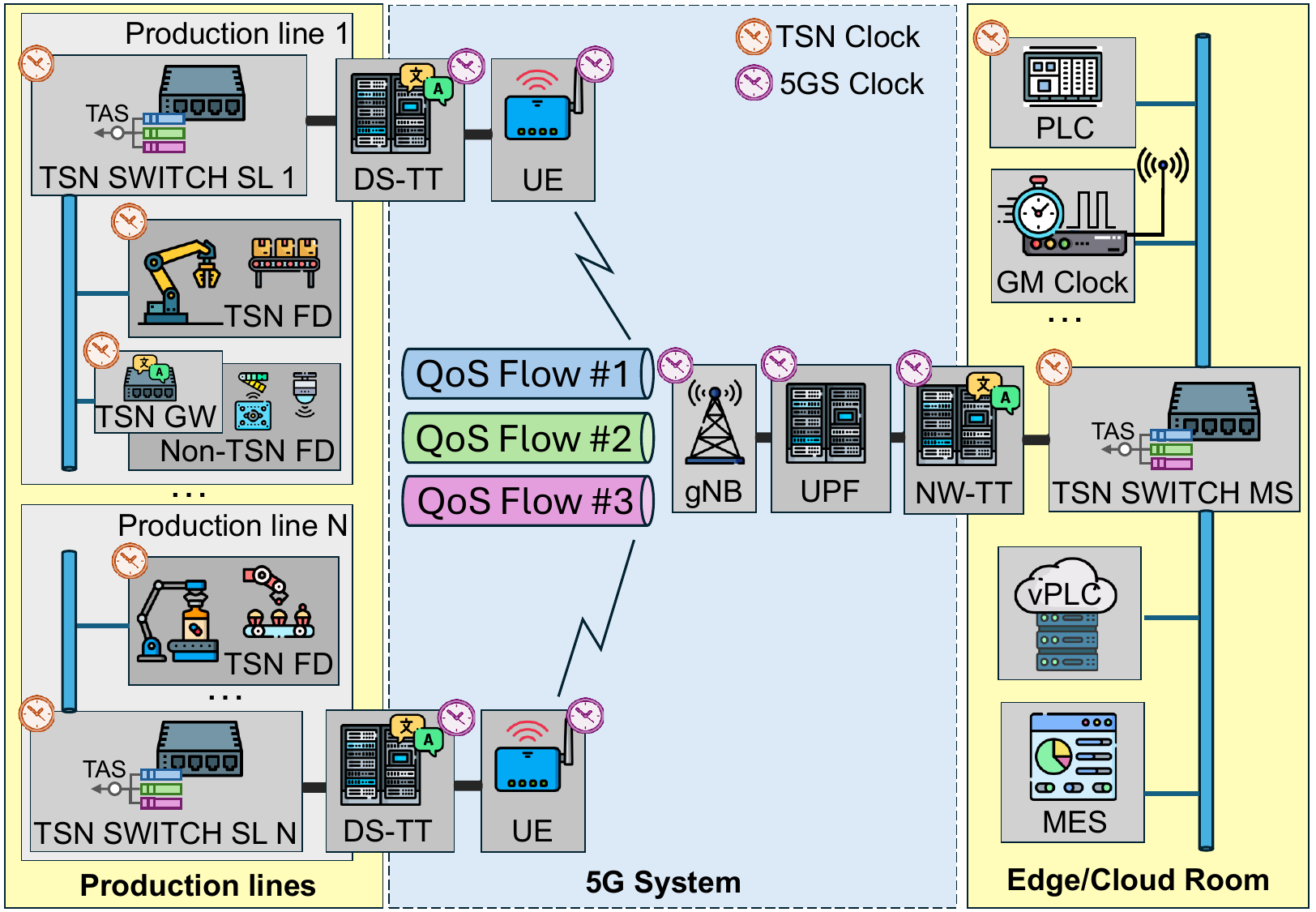}
    \caption{\gls{5G}-\gls{TSN} network architecture in an Industry 4.0 factory.}
   \label{fig:architecture}
\end{figure}

As depicted in Fig. \ref{fig:architecture}, three connectivity segments are defined in a \gls{5G}-\gls{TSN}-based industrial network~\cite{5GACIA-whitepaperI}:

\begin{itemize}
     \item \textit{\textbf{Edge/Cloud Room}}: Centralizes management tasks handled by the \gls{MES}, such as monitoring, data collection, and analytics. Control functions are traditionally performed by \glspl{PLC}, which may run on dedicated hardware or general-purpose servers, i.e., \glspl{vPLC}. This layer may also include a network device that provides the \gls{TSN} \gls{GM} clock reference, typically derived from \gls{GNSS}~\cite{Darroudi2024}, for distribution across the network. 
     
    \item \textit{\textbf{\gls{5G} System}}: According to \gls{3GPP} TS 23.501 (v19.0.0)~\cite{3gpp_ts_23_501_v19_0_0}, the \gls{5G} system integrates into the \gls{TSN} network as one or more virtual \gls{TSN} switches, with the \glspl{UPF} and \glspl{UE} acting as endpoints. The \gls{UE} connects wirelessly to the \gls{gNB}. The \gls{TSN} Translators, specifically the \gls{NW-TT} located in the \gls{UPF} and the \gls{DS-TT} in the \gls{UE}, support the integration between the \gls{TSN} and \gls{5G} domains by adapting traffic formats and \gls{QoS} information, and enabling the transport of synchronization information.
    
    \item \textit{\textbf{Production Lines}}: Each includes \glspl{FD} such as sensors and actuators, along with local \glspl{PLC} for distributed control. \glspl{FD} report operational data to centralized \glspl{PLC}, enabling hierarchical decision-making. Each production line connects to a \gls{TSN} \gls{SL} switch that receives clock signals from the \gls{TSN} \gls{MS} switch via the \gls{5G} system and redistributes synchronization to the \glspl{FD} within this production line.
\end{itemize}

\subsection{Industrial applications}\label{sec:industrial_applications}
Industrial network traffic is predominantly delay-sensitive, with \gls{E2E} latency requirements ranging from hundreds of microseconds to few tens of milliseconds~\cite{Ojcom_Oscar}. Although other traffic types exist, such as network control, mobile robotics, and video streams, 
\gls{TAS} can be applied to \textit{Cyclic-Synchronous} applications, which require highly predictable timing to ensure reliable communications~\cite{5GACIA-whitepaperI,IIC_traffic}. 

The \textit{Cyclic-Synchronous} applications consist of periodic communication between devices operating on independent cycles, with synchronization enforced at intermediate network nodes rather than end devices. Each device samples and updates at its own rate, allowing for bounded jitter and some timing variation. Although the \gls{E2E} packet transmission delay must remain within predictable bounds, occasional variation is tolerated. Thereby, jitter is constrained to the latency bound \cite{IIC_traffic}. This traffic is commonly used in controller-to-I/O exchanges, periodic sensor polling, and updates to supervisory systems. Examples include \gls{PLC}-to-actuator response commands, graphic updates to \gls{SCADA} systems, and routine diagnostic or historian data transfers.

In addition, another category of time-sensitive industrial applications coexists with \textit{Cyclic-Synchronous}: the \textit{Isochronous}. Although both of them require strict delay and jitter analysis in \gls{5G}-\gls{TSN} networks, our work addresses general \textit{Cyclic-Synchronous} applications and evaluates the feasibility of their scheduling, as the stringent requirements of \textit{Isochronous} applications cannot currently be met, which significantly exceed the latency capabilities of existing \gls{5G} deployments~\cite{Damsgaard2023}.

\subsection{QoS Traffic Management}
Traffic prioritization in \gls{TSN} networks relies on the 3-bit \gls{PCP} field defined in IEEE 802.1Q \gls{VLAN} tags, allowing up to eight priority levels~\cite{ieee8021q}. These levels enable differentiation according to \gls{QoS} requirements: higher values (i.e., \gls{PCP} 4–7) are typically assigned to critical traffic, while lower ones (i.e., \gls{PCP} 0–3) serve less time-sensitive or best-effort data~\cite{5GACIA-whitepaperI}.

In \gls{5G} networks, \gls{QoS} is managed for each flow by a \gls{QFI} and associated with a standardized \gls{5QI}, as specified in \gls{3GPP} TS 23.501~\cite{3gpp_ts_23_501_v19_0_0}. Each \gls{5QI} defines key performance characteristics such as priority level, delay tolerance, and packet error rate, which determine the treatment of traffic throughout the \gls{5G} system~\cite{Perdigao2024}.

While \gls{TSN} enforces \gls{QoS} through \gls{PCP}-based prioritization, \gls{5G} employs \gls{5QI}-driven flow control to differentiate traffic. The mapping between \gls{TSN} traffic classes and \gls{5G} \gls{QoS} flows remains an active research topic, primarily due to the semantic differences between the \gls{PCP}-based prioritization in \gls{TSN} and the \gls{5QI}-based framework in \gls{5G}. As shown in~\cite{Ojcom_Oscar}, a feasible approach involves classifying Ethernet frames based on their \gls{PCP} field at the \gls{UE} and \gls{UPF}, using packet filters to associate them with appropriate \gls{5G} \gls{QoS} flows.

\subsection{IEEE 802.1Qbv Time-Aware Shaper (TAS)}\label{sec:TAS}
IEEE 802.1Qbv is a \gls{TSN} standard that specifies the \gls{TAS} mechanism, which enables time-aware scheduling of Layer~2 frames at the egress ports of \gls{TSN} switches based on \gls{QoS} requirements~\cite{Oge2020,Oliver2018,Walrand2023}. \gls{TAS} utilizes the \gls{PCP} field in the IEEE 802.1Q header to classify packets into one of eight \gls{FIFO} queues. At each egress port, these queues are prioritized to ensure that higher-priority traffic is transmitted before lower-priority traffic.

Each egress port is controlled by a \gls{GCL}, which defines a time-triggered transmission schedule divided into transmission windows governed by the clock reference. During each transmission window, one or more queues are permitted to transmit, depending on the binary state of their associated gates. Each queue has its own gate, and the \gls{GCL} specifies the time intervals during which each gate is open or closed. When multiple gates are open, transmission order typically follows queue priority, although exact behavior may depend on the switch implementation.

\gls{TAS} scheduling is organized around periodic network cycles that enable deterministic communication. A network cycle consists of a fixed-duration time interval which encompasses a full instance of a specific set of transmission windows defined by the \gls{GCL}~\cite{Lin22}. The duration of the network cycle is typically chosen to align with the application cycles involved, which are defined as the periods at which message exchanges occur. This alignment is commonly achieved by selecting the network cycle duration as the greatest common divisor of the involved application cycles. For more information on \gls{TAS} see~\cite{ieee8021qbv}.

\subsection{Synchronization in \gls{5G}-\gls{TSN} Networks}\label{sec:sync}
Time synchronization is essential in \gls{5G}-\gls{TSN} networks to support the deterministic requirements of \gls{IIoT} applications. In typical \gls{TSN} architectures, a \gls{TSN} \gls{MS} switch distributes the \gls{GM} clock via \gls{PTP} or \gls{gPTP} messages to multiple \gls{TSN} \gls{SL} switches, each deployed along a different production line, as defined in Section~\ref{sec:network_segments}. Upon receiving these messages, each \gls{TSN} \gls{SL} switch estimates the time difference between its local clock and the reference clock of the \gls{TSN} \gls{MS} switch, known as the clock offset, and adjusts its local time accordingly~\cite{ieee8021qas}.

According to the architecture defined in \gls{3GPP} TS 23.501~\cite{3gpp_ts_23_501_v19_0_0}, \gls{TSN} translators, specifically the \gls{NW-TT} and \gls{DS-TT}, enable propagation of the \gls{GM} clock across the \gls{5G} system to the \gls{TSN} domain, thus maintaining clock consistency across \gls{TSN} switches interconnected via \gls{5G} (see Fig.~\ref{fig:architecture}). A widely adopted configuration for propagating synchronization over \gls{5G} is the \gls{TC} mode defined in IEEE 1588 ~\cite{StandadsIEEE1588_2019,striffler2021,caleya2025empirical}, where synchronization messages are forwarded with the \texttt{correctionField} updated to reflect the residence time within each intermediate node, while original timestamps remain unchanged. Unlike \gls{BC} mode, where each node terminates and regenerates synchronization messages, \gls{TC} mode preserves a single timing domain by accumulating residence times~\cite{ munoz2025}. The \gls{NW-TT} and \gls{DS-TT} measure the residence time within the \gls{5G} system and include this delay in the forwarded messages with the \texttt{correctionField}. This operation complies with IEEE 1588-2019 and enables accurate clock correction at the \gls{TSN} endpoint.  For more information see \cite{ieee8021qas, munoz2025,StandadsIEEE1588_2019,caleya2025empirical,striffler2021}.

Discrepancies in the clocks of different devices within the \gls{5G}-\gls{TSN} network may occur, preventing the devices from updating their clocks accurately. The \gls{3GPP} TS 22.104 \cite{3gpp_ts22104} specifies that a maximum clock drift contribution of 900 ns must be guaranteed for \gls{5G} systems to enable time-critical industrial applications. In line with this, the work in \cite{caleya2025empirical} empirically quantizes a maximum peak-to-peak synchronization error of 500 ns, which is significantly below the requirement.

\section{System Model}\label{sec:SystemModel}
This section introduces the network and traffic models. We then describe the \gls{TAS} model, followed by a description of the different sources of latency in the system. Table~\ref{table:MainMathNotations} provides a summary of key mathematical notations used throughout the paper.

\begin{table*}[t!]
\centering
\caption{Main Mathematical Notations}\label{table:MainMathNotations}
\resizebox{\textwidth}{!}{
\begin{tabular}{|c|c|c|c|}
\hline
\textbf{Variable} & \textbf{Description} & \textbf{Variable} & \textbf{Description} \\
\hline
\raisebox{-0.3ex}{$\mathcal{I}$} & Set of network nodes. & \raisebox{-0.2ex}{$\mathcal{E}$} & Set of links.\\
\hline
\raisebox{0.2ex}{$\varepsilon_{i,j}$} & Link $\varepsilon$ between network nodes $i$ and $j$, $\forall i,j \in \mathcal{I}$. & \raisebox{-0.4ex}{$\mathcal{I}^{\text{\gls{5G}}}$} & Subset of \gls{5G} network nodes. \\
\hline
\raisebox{-0.5ex}{$\mathcal{E}^{\text{\gls{5G}}}$} & Subset of links within the \gls{5G} system. & \raisebox{-0.4ex}{$\mathcal{I}^{\text{\gls{TSN}}}$} & Subset of \gls{TSN} nodes. \\
\hline
\raisebox{-0.5ex}{$\mathcal{E}^{\text{\gls{TSN}}}$} & Subset of links within the \gls{TSN} system. & $\mathcal{S}$ & Set of traffic flows. \\ 
\hline
$s$ & Traffic flow type $s \in \mathcal{S}$. & \raisebox{-0.3ex}{$T_s^{\text{app}}$} & \textit{Application cycle} for traffic flow $s \in \mathcal{S}$. \\ 
\hline
$N_s$ & Packets generated in an \textit{application cycle} for traffic flow $s \in \mathcal{S}$. & $ L_s$ & Packet size for traffic flow $s \in \mathcal{S}$. \\ 
\hline
\raisebox{-0.2ex}{$R_s^{\text{gen}}$} & Data rate for traffic flow $s \in \mathcal{S}$. & \raisebox{-0.2ex}{$d_s^{\text{\gls{E2E}}}$} & \gls{E2E} packet transmission delay for traffic flow $s \in \mathcal{S}$. \\ 
\hline
$D_s$ & Delay bound for \gls{E2E} transmission time for traffic flow $s \in \mathcal{S}$. & $Q_i$ & Set of queues in an output port in node $i \in \mathcal{I}^{\text{\gls{TSN}}}$. \\ 
\hline
$q$ & Output port queue $q \in \mathcal{Q}_i$. & $G_{i,q}(t)$ & Binary function to open/close the gate for queue  $q \in \mathcal{Q}_i$ at node $i \in \mathcal{I}^{\text{\gls{TSN}}}$. \\ 
\hline
$T_i^{\text{nc}}$ & \textit{Network cycle} in node $i \in \mathcal{I}^{\text{\gls{TSN}}}$. & $T_{i,q}^{\text{open}} $ & Time instant the gate is open for queue $q \in \mathcal{Q}_i$ in node $i \in \mathcal{I}^{\text{\gls{TSN}}}$. \\ 
\hline
$T_{i,q}^{\text{closed}}$ & Time instant the gate is closed for queue  $q \in \mathcal{Q}_i$ in node $i \in \mathcal{I}^{\text{\gls{TSN}}}$. & $ W_{i,s} $ & \textit{Transmission windows} in node $i \in \mathcal{I}^{\text{\gls{TSN}}}$ for traffic flow $s \in \mathcal{S}$. \\ 
\hline
$T^{\text{GB}}$ & Duration of the guard band. & $d_i^{\text{que,in}}$ & Packet waiting time in the input queue in node $i \in \mathcal{I}$. \\ 
\hline
$d_i^{\text{proc}}$ & Packet processing time in node $i \in \mathcal{I}$. & $d_{i,q}^{\text{que,out}}$ & Packet waiting time in the output queue $q \in \mathcal{Q}_i$ in node $i \in \mathcal{I}$. \\ 
\hline
$d_{\varepsilon_{i,j},s}^{\text{tran}}$ & Packet transmission time at link $\varepsilon_{i,j} \in \mathcal{E}$ for traffic flow $s \in \mathcal{S}$. & $r_{\varepsilon_{i,j}}$ & Data rate at link $\varepsilon_{i,j} \in \mathcal{E}$. \\ 
\hline
$D_{\varepsilon_{i,j}}^{\text{prop}} $ & Propagation time in the physical medium of link $\varepsilon_{i,j} \in \mathcal{E}$. & \raisebox{-0.3ex}{$d_s^{\text{\gls{5G}}}$} & Packet transmission time within the \gls{5G} system for traffic flow $s \in \mathcal{S}$. \\ 
\hline
\raisebox{-0.5ex}{$d_s^{\text{\gls{MS},\gls{SL}}}$} & Packet delay between \gls{MS} and \gls{SL} output ports for traffic flow $s \in \mathcal{S}$. & \raisebox{-0.5ex}{$\widetilde{d}_s^{\text{\gls{MS},\gls{SL}}}$} & Packet delay from \gls{MS} output port to \gls{SL} processing for traffic flow $s \in \mathcal{S}$. \\ 
\hline
\raisebox{0.1ex}{$\Delta_{i,j}$} & Synchronization error between \gls{TSN} nodes $i,j \in \mathcal{I^{\text{\gls{TSN}}}}$. & \raisebox{-0.3ex}{$d_s^{\text{emp}}$} & Empirical delay measured in our study for traffic flow $s \in \mathcal{S}$. \\ 
\hline
\raisebox{-0.5ex}{$\widetilde{d}_s^{\text{emp}}$} & ZWSL empirical delay measured in our study for traffic flow $s \in \mathcal{S}$. & $\delta_s$ & \textit{Offset} between two adjacent \gls{TSN} switches for traffic flow $s \in \mathcal{S}$. \\ 
\hline
$\hat{D}_{s,p}^{\text{emp}} $ & Percentile $p$ of the ZWSL empirical delay distribution for traffic flow $s\in\mathcal{S}$. & $\delta'_s$ & \textit{Network cycle offset} for traffic flow $s \in \mathcal{S}$. \\ 
\hline
 $ t_s^{\text{uni}}$ & \textit{Uncertainty interval} for traffic flow $s\in\mathcal{S}$. & \raisebox{-0.5ex}{$t_s^{\text{jit}}$} & \gls{5G}-\gls{TSN} network jitter for traffic flow $s\in\mathcal{S}$.  \\
 \hline
\end{tabular}
}
\end{table*}

\textbf{Notation Conventions.} We use calligraphic letters (e.g.,~$\mathcal{X}$) to denote sets. Lowercase letters (e.g.,~$y$) represent random variables, while uppercase letters (e.g.,~$Y$) denote constant parameters. Binary variables are typeset in uppercase sans serif font (e.g.,~$\mathsf{X}$). Subscripts indicate that a parameter applies to specific elements of a given set; for example, $z_{i,j}$ refers to the parameter $z$ corresponding to elements $i \in \mathcal{I}$ and $j \in \mathcal{J}$. Superscripts provide descriptive annotations, e.g.,~$z^{\text{desc}}$ denotes the variable $z$ with descriptor "desc". In addition, $f_x(\cdot)$ and $F_x(\cdot)$ denote the \gls{PDF} and \gls{CDF} of the random variable $x$, respectively. Finally, the letter $\hat{Z}$ denotes the statistical upper bound of $F_x(\cdot)$.

\subsection{Network Model} \label{sec:network_model}
We consider a set of network nodes denoted by $\mathcal{I}$, comprising: (\textit{i}) two \gls{TSN} switches, denoted as master switch \gls{MS} and slave switch \gls{SL}, respectively; (\textit{ii}) two \gls{TSN} translators, one being a network-side translator and denoted as \gls{NW-TT} and the other being the device-side translator and denoted as \gls{DS-TT}; (\textit{iii}) a \gls{5G} \gls{UE} denoted as \gls{UE}; and (\textit{iv}) a \gls{5G} \gls{gNB} and an \gls{UPF}, denoted by \gls{gNB} and \gls{UPF}, respectively. Each communication link is represented by $\varepsilon$, and the set of all such links is denoted by $\mathcal{E}$. A specific link between nodes $i$ and $j$ is denoted by $\varepsilon_{i,j} \in \mathcal{E}$, where $i, j \in \mathcal{I}$. The topology is defined by the sequential links: $\mathcal{E} \equiv \{\varepsilon_{\text{\gls{MS}},\text{\gls{NW-TT}}}, \varepsilon_{\text{\gls{NW-TT}},\text{\gls{UPF}}}, \varepsilon_{\text{\gls{UPF}},\text{\gls{gNB}}}, \varepsilon_{\text{\gls{gNB}},\text{\gls{UE}}}, \varepsilon_{\text{\gls{UE}},\text{\gls{DS-TT}}}, \varepsilon_{\text{\gls{DS-TT}},\text{\gls{SL}}}\}$. We define the subset of nodes corresponding to the \gls{5G} system as $\mathcal{I}^{\text{\gls{5G}}} \equiv \{\text{\gls{UE}}, \text{\gls{gNB}}, \text{\gls{UPF}}, \text{\gls{NW-TT}}, \text{\gls{DS-TT}}\}$. Similarly, the virtual \gls{5G} system link set $\mathcal{E}^{\text{\gls{5G}}} \subset \mathcal{E}$ contains the subset of physical links that connect the nodes of the \gls{5G} system, i.e., $\mathcal{E}^{\text{\gls{5G}}} \equiv \{\varepsilon_{\text{\gls{NW-TT}}, \text{\gls{UPF}}}, \varepsilon_{\text{\gls{UPF}}, \text{\gls{gNB}}}, \varepsilon_{\text{\gls{gNB}}, \text{\gls{UE}}}, \varepsilon_{\text{\gls{UE}}, \text{\gls{DS-TT}}}\}$.  Finally, we define the subset of \gls{TSN} switches as $\mathcal{I}^{\text{\gls{TSN}}} \subset \mathcal{I}$, i.e., $\mathcal{I}^{\text{\gls{TSN}}} \equiv \{\text{\gls{MS}}, \text{\gls{SL}}\}$, which are interconnected via the \gls{5G} system with the link set $\mathcal{E}^{\text{\gls{TSN}}} \subset \mathcal{E}$, containing the subset of physical links to the \gls{5G} bridge bounds \gls{NW-TT} and \gls{DS-TT}, i.e., $\mathcal{E}^{\text{\gls{TSN}}} \equiv \{\varepsilon_{\text{\gls{MS},\gls{NW-TT}}}, \varepsilon_{\text{\gls{DS-TT},\gls{SL}}}\}$, respectively.

\subsection{Traffic Model and QoS Level Assignment} \label{sec:traffic_model}
Let $\mathcal{S}$ denote the set of traffic flows traversing the considered \gls{5G}-\gls{TSN} network.  Specifically, $\mathcal{S}$ includes:

\begin{itemize}
    \item \textbf{\textit{A downlink \gls{DC} flow}} generated by a \textit{Cyclic-Synchronous} application, as described in Section~\ref{sec:industrial_applications}. We assume a \gls{DC} flow in downlink as a set of packets sharing a source at the Edge/Cloud Room, e.g., a \gls{PLC}, and any of the devices in the same production line as the destination, e.g., actuators, which are typically served by a common switch, i.e., the \gls{SL} switch. Each \textit{application cycle}, of periodic duration $T_{\text{\gls{DC}}}^{\text{app}}$, the \gls{PLC} generates a batch of $N_{\text{\gls{DC}}}$ packets of constant size $L_{\text{\gls{DC}}}$, resulting in an average data rate $R_{\text{\gls{DC}}}^{\text{gen}}~=~N_{\text{\gls{DC}}}~\cdot~L_{\text{\gls{DC}}} / T_{\text{\gls{DC}}}^{\text{app}}$, as a response delivered to all these actuators after processing the production state \cite{belliardi2018use}. Additionally, packets must traverse the \gls{5G}-\gls{TSN} network subject to an \gls{E2E} delay constraint $d_{\text{\gls{DC}}}^{\text{\gls{E2E}}}~\leq~D_{\text{\gls{DC}}}$. Assuming these packets belong to a single application, they share the same timing constraints between them.
    
    \item \textbf{\textit{A downlink \gls{BE} flow}} composed of packets that do not require strict timing guarantees. We assume packets of constant size  $L_{\text{\gls{BE}}}$ are generated at a constant data rate $R_{\text{\gls{BE}}}^{\text{gen}}$.
    
    \item \textbf{\textit{Uplink and downlink \gls{PTP} flows}} are considered to support clock synchronization among \gls{TSN} switches. The exchange of these messages, as defined by the \gls{PTP} standard, occurs periodically, with an \textit{application cycle} $T_{\text{\gls{PTP}}}^{\text{app}}$ significantly larger than $T_{\text{\gls{DC}}}^{\text{app}}$, i.e., $T_{\text{\gls{PTP}}}^{\text{app}} \gg T_{\text{\gls{DC}}}^{\text{app}}$. 
\end{itemize}

\gls{PTP} flows are assigned the highest priority, followed by \gls{DC} and then \gls{BE} flows, consistently across both the \gls{5G} and \gls{TSN} domains. Accordingly, \gls{PTP} and \gls{DC} packets are assigned higher \gls{PCP} values for \gls{TSN} scheduling, and they are mapped to \gls{5G} \gls{QoS} flows with lower \gls{5QI} indices, indicating stricter \gls{QoS} treatment. \gls{BE} packets are mapped to the lowest priority class with higher \gls{5QI} index.

\subsection{TAS model}\label{sec:TAS_model} 
At each \gls{TSN} switch $i \in \mathcal{I}^{\text{\gls{TSN}}}$, each egress port is associated with a set $\mathcal{Q}_i$ containing up to eight output queues. We assume a one-to-one mapping between each queue $q \in \mathcal{Q}_i$ and a traffic flow $s \in \mathcal{S}$, allowing interchangeability of $q$ and $s$ throughout this paper. Furthermore, we assume the \gls{GCL} enforces mutually exclusive gate openings among the eight queues per egress port, guaranteeing that only one queue is permitted to transmit at any given instant.

Accordingly, the \gls{GCL} configuration for queue $q \in \mathcal{Q}_i$ at switch $i \in \mathcal{I}^{\text{\gls{TSN}}}$ is formally expressed by Eq.~\eqref{eqn:gcl_open}. The binary variable $\mathsf{G}_{i,q}(t)$ indicates whether the gate is open (1) or closed (0). The gates operate periodically with period $T_i^{\text{nc}}$, referred to as the \textit{network cycle}. In the \textit{network cycle} $n=0$, the gate opening and closing instants, $T_{i,q}^{\text{open}}$ and $T_{i,q}^{\text{closed}}$, define the \textit{transmission window}  duration $W_{i,q} = T_{i,q}^{\text{closed}} - T_{i,q}^{\text{open}}$.
\begin{equation}
\mathsf{G}_{i,q}(t) =
\begin{cases} 
    1, &  nT_i^{\text{nc}}+T_{i,q}^{\text{open}} < t \leq  nT_i^{\text{nc}}+T_{i,q}^{\text{closed}},\\
     &  \; \forall n  \in \mathbb{N} \cup \{0\}. \\
    0, &  \text{otherwise}.
\end{cases}  
\label{eqn:gcl_open}
\end{equation}

The \gls{DC} flow period $T_{\text{\gls{DC}}}^{\text{app}}$ typically ranges from several hundreds of microseconds up to a few tens of milliseconds, whereas \gls{PTP} synchronization messages are generated approximately every $T_{\text{\gls{PTP}}}^{\text{app}} \approx 1\,\text{s}$. Given that condition $T_{\text{\gls{PTP}}}^{\text{app}}~\gg~T_{\text{\gls{DC}}}^{\text{app}}$, we consider the duration of the \textit{network cycle}, $T_i^{\text{nc}}=T_{\text{\gls{DC}}}^{\text{app}}$, $\forall i \in \mathcal{I}^{\text{\gls{TSN}}}$, as the \gls{DC} flow is the primary target of this work. 

\begin{figure}[b!]
    \centering
   \includegraphics[width=\columnwidth]{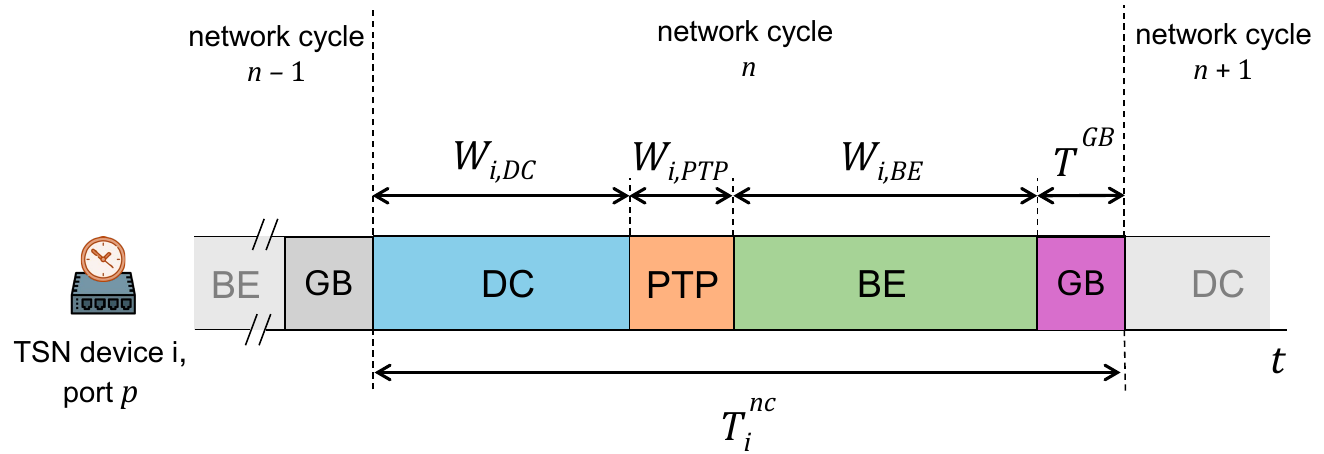}
    \caption{Diagram of \gls{TAS}-scheduled network cycles in \gls{GCL}.}
   \label{fig:network_cycle}
\end{figure}

Each \textit{network cycle} comprises three non-overlapping \textit{transmission windows} (see Fig. \ref{fig:network_cycle}): $W_{i,\text{\gls{DC}}}$ for the \gls{DC} traffic, $W_{i,\text{\gls{PTP}}}$ for \gls{PTP} synchronization messages and $W_{i,\text{\gls{BE}}}$ for \gls{BE} traffic, followed by a fixed guard band $T^{\text{GB}}$ to avoid interference on \gls{DC} traffic. Thus, $T_i^{\text{nc}}~=~W_{i,\text{\gls{DC}}}~+~W_{i,\text{\gls{PTP}}}~+~W_{i,\text{\gls{BE}}}~+~T^{\text{GB}}$, $\forall~i~\in~\mathcal{I}^{\text{\gls{TSN}}}$. We consider $W_{i,\text{\gls{PTP}}}$ occupies a negligible fraction of $T_i^{\text{nc}}$, 100 to 1000 times smaller, due to the low frequency of synchronization messages, i.e., $W_{i,\text{\gls{PTP}}} \ll W_{i,\text{\gls{DC}}}$ and $W_{i,\text{\gls{PTP}}} \ll W_{i,\text{\gls{BE}}}$. Due to this and for ease of reading, $W_{i,\text{\gls{PTP}}}$ is omitted in subsequent equations, while it is implicitly assumed to be scheduled immediately after $W_{i,\text{\gls{DC}}}$. For further details on \gls{PTP} planning, see~\cite{munoz2025}.

Finally, at the \gls{MS}, each batch of $N_{\text{\gls{DC}}}$ packets belonging to the \gls{DC} flow is assumed to be available at the start of every \textit{network cycle}.

\subsection{Delay Model}
\begin{figure*}[t!]
\centering
\includegraphics[width=\textwidth]{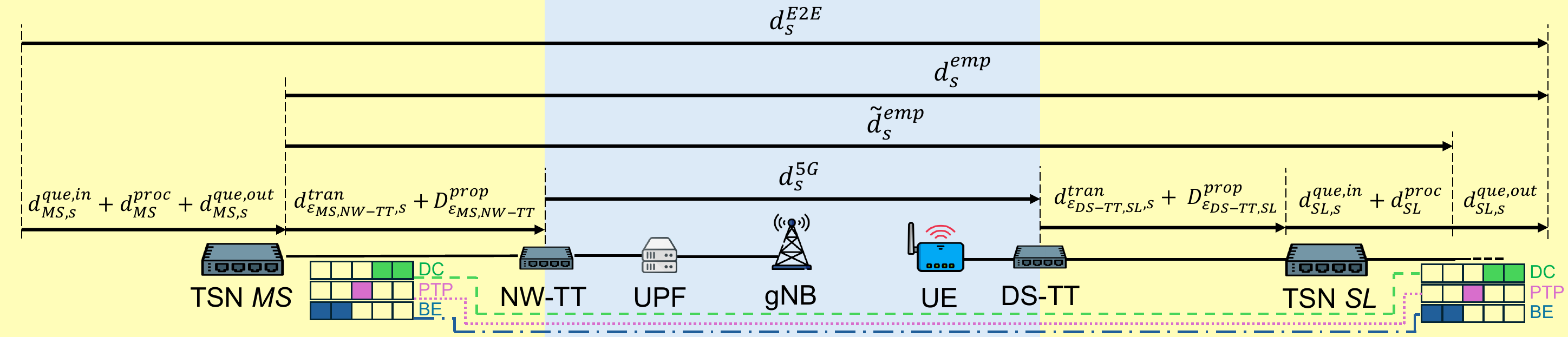}
\caption{\gls{5G}-\gls{TSN} network elements and associated delay terms in the downlink transmission for each traffic flow $s\in \mathcal{S}$.}
\label{fig:network_model}
\end{figure*}

For each packet of flow $s \in \mathcal{S}$ traversing node $i \in \mathcal{I}$, the total delay comprises five components: input queuing delay, processing delay, output queuing delay, transmission delay, and propagation delay. These can be seen in Fig.~\ref{fig:network_model}.

The input queuing delay $d_{i,s}^{\text{que,in}}$ is the interval between the arrival of a packet at node $i$ and the start of its processing. The processing delay $d_{i,s}^{\text{proc}}$ corresponds to the time required by node $i$ to parse the packet header and determine the forwarding action. The output queuing delay $d_{i,s}^{\text{que,out}}$ refers to the delay from the end of processing at node $i$ until the packet is transmitted to the next hop. 

The transmission delay $d_{\varepsilon_{i,j},s}^{\text{tran}}$ corresponds to the time needed to serialize all bits of the packet over the link $\varepsilon_{i,j} \in \mathcal{E}$. It depends on the packet size $L_s$ and the link capacity $r_{\varepsilon_{i,j}}$, and is given by $
d_{\varepsilon_{i,j},s}^{\text{tran}} = L_s/r_{\varepsilon_{i,j}}$. The propagation delay $D_{\varepsilon_{i,j}}^{\text{prop}}$ is the time a signal takes to travel through link $\varepsilon_{i,j} \in \mathcal{E}$ and is assumed to be constant for any flow $s \in \mathcal{S}$ in that link in time.

\section{Analysis of \gls{5G} Delay and Jitter on \gls{TAS} scheduling}\label{sec:jitter}
In this section, we analyze how \gls{5G} impacts \gls{TAS} scheduling performance. First, we define the \gls{E2E} delay for a packet of an arbitrary flow and study how the \gls{5G} delay component impacts it. Then, we formalize the constraint for the \textit{transmission window} of the \gls{DC} flow in \gls{TSN} switches. Next, we introduce the concept of \textit{offset} between the \textit{network cycles} of the \gls{MS} and \gls{SL} switches, and derive the conditions required to ensure deterministic communication. Finally, we analyze how this \textit{offset} interacts with the \gls{5G} delay component, and evaluate how different \gls{TAS} parameter configurations influence the scheduling performance across various scenarios.

\subsection{Analysis of Delay Components}\label{sec:DelayComponents}
The set of flows $\mathcal{S}$ traverses a sequence of network nodes to reach their destination, as illustrated in Fig.~\ref{fig:network_model}.  The \gls{E2E} packet transmission delay $d_{s}^{\text{\gls{E2E}}}$ for the flow $s \in \mathcal{S}$ along this path is computed as the sum of delays at each network node plus the transmission delays on each link:
\begin{equation}
d_{s}^{\text{\gls{E2E}}}\hspace{-0.05cm}=\hspace{-0.1cm}\sum_{i \in \mathcal{I}} \left( d_{i,s}^{\text{que,in}}\hspace{-0.1cm}+\hspace{-0.05cm}d_{i,s}^{\text{proc}}\hspace{-0.05cm}+\hspace{-0.05cm}d_{i,s}^{\text{que,out}} \right)\hspace{-0.05cm}+\sum_{e \in \mathcal{E}}\left( d_{e,s}^{\text{tran}}\hspace{-0.1cm}+\hspace{-0.05cm}D_{e}^{\text{prop}} \right).
\label{eqn:e2e_delay}
\end{equation}
\indent The \gls{5G} delay $d_s^{\text{\gls{5G}}}$ is defined as the sum of node processing and queuing delays, plus link transmission in the \gls{5G} domain:
\begin{equation}
d_{s}^{\text{\gls{5G}}}=\hspace{-0.05cm}\sum_{j \in \mathcal{I^{\text{\gls{5G}}}}}\hspace{-0.1cm}\left( d_{j,s}^{\text{que,in}}\hspace{-0.05cm}+\hspace{-0.05cm}d_{j,s}^{\text{proc}}\hspace{-0.05cm}+\hspace{-0.05cm}d_{j,s}^{\text{que,out}} \right)\hspace{-0.03cm}+\hspace{-0.1cm}\sum_{k \in \mathcal{E^{\text{\gls{5G}}}}}\hspace{-0.1cm}\left( d_{k,s}^{\text{tran}}\hspace{-0.05cm}+\hspace{-0.05cm}D_{k}^{\text{prop}} \right).
\label{eqn:5G_delay}
\end{equation}
\indent To assess the impact of \gls{5G} in combination with the \gls{TAS} scheduling, we consider the packet delay between the \gls{MS} and \gls{SL} output ports, $d_{s}^{\text{\gls{MS},\gls{SL}}}$, and compute it as follows:
\begin{equation}
d_{s}^{\text{\gls{MS},\gls{SL}}}=\hspace{-0.1cm}\sum_{\varepsilon \subset \mathcal{E}^{\text{\gls{TSN}}}}\hspace{-0.2cm}\left(d_{\varepsilon,s}^{\text{tran}}\hspace{-0.cm}+\hspace{-0.05cm}D_{\varepsilon}^{\text{prop}}\hspace{-0.1cm}\hspace{0.1cm}\right)+d_{s}^{\text{\gls{5G}}}+d_{\text{\gls{SL}},s}^{\text{que,in}} +
d_{\text{\gls{SL}}}^{\text{proc}} + d_{\text{\gls{SL}},s}^{\text{que,out}} \hspace{-0.05cm}.
\label{eqn:emp_delay_noSync}
\end{equation}
\indent For convenience, we also consider the  packet delay between the \gls{MS} output port and the \gls{SL} switch processing, $\widetilde{d}_{s}^{\text{\gls{MS},\gls{SL}}}$, that is, excluding the \gls{SL} output queuing delay, $d_{\text{\gls{SL}},s}^{\text{que,out}}$, as it is influenced by the \gls{TAS} configuration:
\begin{equation}
\widetilde{d}_{s}^{\text{\gls{MS},\gls{SL}}}=\sum_{\varepsilon \subset \mathcal{E}^{\text{\gls{TSN}}}}\left(d_{\varepsilon,s}^{\text{tran}}\hspace{-0.05cm}+D_{\varepsilon}^{\text{prop}}\right)+d_{s}^{\text{\gls{5G}}}+d_{\text{\gls{SL}},s}^{\text{que,in}} + d_{\text{\gls{SL}}}^{\text{proc}}\hspace{-0.05cm}.
\label{eqn:emp_delay_noSync_noQueue}
\end{equation}
\indent In this work, we rely on empirical delay measurements for our analysis. Therefore, our model has to consider the synchronization error that inherently affects the delay measurement between separate \gls{TSN} nodes, i.e., $\Delta_{i,j}$, $\forall i,j \in \mathcal{I^{\text{\gls{TSN}}}}$. Hence, the empirical measurement of the packet delay between the \gls{MS} and the \gls{SL} output ports, $d_{s}^{\text{emp}}$, could be expressed as in Eq.~\eqref{eqn:emp_delay}, where $\Delta_{\text{\gls{MS},\gls{SL}}}$ refers to the synchronization error between \gls{MS} and \gls{SL}.
\begin{equation}
d_{s}^{\text{emp}} =d_{s}^{\text{\gls{MS},\gls{SL}}} + \Delta_{\text{\gls{MS},\gls{SL}}}.
\label{eqn:emp_delay}
\end{equation}
\indent The value of  $\Delta_{\text{\gls{MS},\gls{SL}}}$ is assumed to take positive or negative values, as clocks may be ahead or behind each other at any instant. A higher $|\Delta_{\text{\gls{MS},\gls{SL}}}|$ may cause the measurements to become unreliable, as it distorts the temporal correspondence between events.
In this way, the \gls{E2E} latency from Eq. \eqref{eqn:e2e_delay} is also affected by the synchronization error. Thus, its empirical measurements can be written as $d_s^{\text{E2Emp}}$ in Eq. \eqref{eqn:e2e_delay2}.
\begin{equation}
d_{s}^{\text{E2Emp}} = d_{\text{\gls{MS}},s}^{\text{que,in}} + d_{\text{\gls{MS}}}^{\text{proc}} + d_{\text{\gls{MS}},s}^{\text{que,out}} + d_{s}^{\text{emp}}.
\label{eqn:e2e_delay2}
\end{equation}
\indent Similarly, the empirical measurement of the packet delay between the \gls{MS} output port and the \gls{SL} switch processing, from now on \gls{ZWSL} empirical delay, $\widetilde{d}_{s}^{\text{emp}}$, could be expressed as in Eq. \eqref{eqn:emp_delay_noQueue}.
\begin{equation}
\widetilde{d}_{s}^{\text{emp}} =\widetilde{d}_{s}^{\text{\gls{MS},\gls{SL}}} + \Delta_{\text{\gls{MS},\gls{SL}}}.
\label{eqn:emp_delay_noQueue}
\end{equation} 
\textbf{\textit{Observation}--.} The \gls{5G} system delay, $d_s^{\text{\gls{5G}}}$, is significantly larger than the transmission delays over wired links, $d_{\varepsilon_{i,j},s}^{\text{tran}}$, $\forall \varepsilon_{i,j} \in \mathcal{E}$\textbackslash$\{\varepsilon_{\text{\gls{gNB},\gls{UE}}}\}$, $\forall i,j \in \mathcal{I}$; the propagation delays, $D_{\varepsilon_{i,j}}^{\text{prop}}$, $\forall \varepsilon_{i,j} \in \mathcal{E}$, $\forall i,j \in \mathcal{I}$; and processing delays in the \gls{TSN} switches,  $d_{i,s}^{\text{proc}}, \forall i \in \mathcal{I^{\text{\gls{TSN}}}}$. On the one hand,  transmission delays over wired links are typically within the microsecond range. For example, a 200~Bytes packet, assuming 42 Bytes of overhead, has a transmission delay of 1.9~$\mu$s in  1~Gbps links. Similarly, processing delays in \gls{TSN} switches are typically also in the microsecond range~\cite{Xue2024}. On the other hand, \gls{5G} system delay is in the range from milliseconds to a few tens of milliseconds~\cite{Damsgaard2023}. This delay and jitter dominance will be corroborated experimentally in Section \ref{sec:Results}. 

As a consequence of this observation, the \gls{5G} system delay, and its associated jitter, play a prominent role in Eq. \eqref{eqn:emp_delay_noSync}-\eqref{eqn:emp_delay_noQueue}, and therefore in the \gls{TAS} configuration of the \gls{TSN} switches.
\indent Since our analysis targets the \gls{DC} flow, the formulation presented from this point onward assumes $ s = \text{\gls{DC}}$.

\subsection{\textit{Transmission Window} Size in \gls{TAS} Scheduling}\label{sec:transmission_windows}
The \textit{transmission window} duration $W_{i,\text{\gls{DC}}}$ for flow $\text{\gls{DC}} \in \mathcal{S}$ at \gls{TSN} switch $i \in \mathcal{I}^{\text{\gls{TSN}}}$ must satisfy two conditions: it must be strictly shorter than the \textit{network cycle} $T_i^{\text{nc}}$ and equal or greater than the cumulative transmission time of $N_{\text{\gls{DC}}}$ packets through the output port. These constraints are formalized in Eq.  \eqref{eqn:window_bounds}, where $j \in \mathcal{I}$ is the next network node after switch $i$.
\begin{equation}
 N_{\text{\gls{DC}}} \cdot d_{\varepsilon_{i,j},\text{\gls{DC}} }^{\text{tran}}\;  \leq W_{i,\text{\gls{DC}}} < T_i^{\text{nc}}.
\label{eqn:window_bounds}
\end{equation}
\indent Violating these bounds can lead to performance degradation. If $W_{i,\text{\gls{DC}}}$ is too short, not all $N_{\text{\gls{DC}} }$ packets can be transmitted within a single \textit{network cycle}. The remaining packets accumulate and are deferred to subsequent \textit{network cycles}, introducing additional delays that are multiples of $T_i^{\text{nc}}$. On the other hand, if $W_{i,\text{\gls{DC}}}$ exceeds the \textit{network cycle} duration, it monopolizes the schedule, preventing other flows $s \in \mathcal{S} \setminus \{\text{\gls{DC}} \}$ from being scheduled during that \textit{network cycle}.

\subsection{TAS Scheduling Offset between \gls{TSN} Switches}\label{sec:offset}
We consider identical \gls{TAS} scheduling configurations at both \gls{MS} and \gls{SL} switches, i.e., $T_{\text{\gls{MS}}}^{\text{nc}} = T_{\text{\gls{SL}}}^{\text{nc}}$ and $W_{\text{\gls{MS}},\text{\gls{DC}}} = W_{\text{\gls{SL}},\text{\gls{DC}}}$. Under this assumption, let us define the \textit{offset} $\delta_{\text{\gls{DC}}}$ as the time difference between the start of the \textit{network cycle} at the \gls{MS} and \gls{SL}. This \textit{offset} must be configured to ensure all $N_{\text{\gls{DC}}}$ packets, generated within a single \textit{application cycle}, arrive at the output queue of the \gls{SL} and are transmitted through its egress port before the corresponding \textit{transmission window} closes. 

Since all $N_{\text{\gls{DC}}}$ packets are sent as a burst from the \gls{MS} into the \gls{5G} system, it is essential to characterize the delay experienced by these packets in the \gls{5G} system to establish a value for the offset $\delta_{\text{\gls{DC}}}$. Assuming that the \gls{5G} system capacity is generally lower than the capacity of a wired link \cite{Damsgaard2023}, the \gls{5G} segment constitutes a bottleneck in the \gls{5G}-\gls{TSN} network, where packets experience increasing queuing delays. The first packet in the burst, if no retransmission is required, may traverse the \gls{5G} network with minimal delay, i.e., $\min(d_{\text{\gls{DC}}}^{\text{\gls{5G}}})$, while each subsequent packet must wait for the transmission of the previous packets. Consequently, delay accumulates across the burst, such that the last packet tends to experience the highest latency, i.e., $ \max(d_{\text{\gls{DC}}}^{\text{\gls{5G}}})$, which already includes the cumulative queuing delay of the entire burst. Consequently, the \textit{offset} $\delta_{\text{\gls{DC}}}$ must be set to at least $\text{max}(\widetilde{d}_{\text{\gls{DC}}}^{\text{emp}})$ to guarantee the availability of packets at the \gls{SL} output port queue to be transmitted on time. This leads to the condition in Eq. \eqref{eqn:offset_over_delay_distribution}.
\begin{equation}
\delta_{\text{\gls{DC}}} \geq \text{max}(\widetilde{d}_{\text{\gls{DC}}}^{\text{emp}}).
\label{eqn:offset_over_delay_distribution}
\end{equation}
\indent Since $\widetilde{d}_{\text{\gls{DC}}}^{\text{emp}}$ is random by nature, it is necessary to characterize its behavior statistically. In this work, we define a statistical upper bound based on a given percentile $p \in [0, 1)$ of the \gls{CDF} $F_{\widetilde{d}_{\text{\gls{DC}}}^{\text{emp}}}(\cdot)$ of the \gls{ZWSL} empirical delay. Specifically, we denote this bound as $\hat{D}_{\text{\gls{DC}},p}^{\text{emp}} = F_{\widetilde{d}_{\text{\gls{DC}}}^{\text{emp}}}^{-1}(p)$, which corresponds to the $p\text{-th}$ percentile of the delay distribution. A higher value of $p$ increases the confidence $\widetilde{d}_{\text{\gls{DC}}}^{\text{emp}}$ will remain below $\hat{D}_{\text{\gls{DC}},p}^{\text{emp}}$~\cite{Egger2025}. For instance, setting $p = 0.999$ yields an upper bound such that 99.9\% of packets experience delays below this value. Accordingly, we set the \textit{offset} $\delta_{\text{\gls{DC}}}$ as in Eq.~\eqref{eqn:upperbound_offset}, ensuring that at least $p \cdot 100$\% of packets have been queued before the \textit{transmission window} in the \gls{SL} closes. 
\begin{equation}
    \delta_{\text{\gls{DC}}} \geq \hat{D}_{\text{\gls{DC},p}}^{\text{emp}}.
    \label{eqn:upperbound_offset}
\end{equation}
\indent An additional parameter of interest is the time instant at which an initial \textit{transmission window} opens at the \gls{SL} for transmitting packets of the \gls{DC} flow. We denote this instant as the \textit{network cycle offset} $\delta_{\text{\gls{DC}}}^{\prime}$, formally defined as follows:
\begin{equation}
\delta_{\text{\gls{DC}}}^{\prime} =
\begin{cases}
\delta_{\text{\gls{DC}}}, & \text{if } \delta_{\text{\gls{DC}}} < 
T_i^{\text{nc}}. \\
\delta_{\text{\gls{DC}}} \bmod
T_i^{\text{nc}}, & \text{otherwise}.
\end{cases}
    \label{eqn:network_cycle_offset}
\end{equation}
\indent The \textit{network cycle offset} $\delta_{\text{\gls{DC}}}^{\prime}$ depends on the configured \textit{offset} $\delta_{\text{\gls{DC}}}$ and the \textit{network cycle} duration $T_i^{\text{nc}}$. When $\delta_{\text{\gls{DC}}}~<~T_i^{\text{nc}}$, it holds that $\delta_{\text{\gls{DC}}}^{\prime}~=~\delta_{\text{\gls{DC}}}$, and the \textit{transmission window} opens exactly at the configured \textit{offset}. Conversely, if $\delta_{\text{\gls{DC}}}~\geq~T_i^{\text{nc}}$, the initial transmission opportunity may occur before the configured \textit{offset}, and the effective opening time is given by $\delta_{\text{\gls{DC}}}^{\prime} = \delta_{\text{\gls{DC}}} \bmod T_i^{\text{nc}} \in [0, T_i^{\text{nc}})$.

\subsection{Conditions for Determinism under 5G Delay and Jitter}\label{sec:determinism_condition}
We consider \textit{deterministic transmission} as the scenario in which the entire burst of $N_{\text{\gls{DC}}}$ packets in the same \textit{transmission window} of a given \textit{network cycle} at the \gls{MS} are delivered and forwarded within a single \textit{transmission window} at the \gls{SL} switch. In this case, the \gls{E2E} jitter remains bounded by the window size $W_{i,\text{\gls{DC}}}$, thus enabling predictable communication. 

To determine if a \textit{deterministic transmission} is possible, it is essential to examine the relationship between the \gls{5G}-induced delay and jitter and the following \gls{TAS} parameters: (i) the \textit{network cycle offset} $\delta_{\text{\gls{DC}}}^{\prime}$, (ii) the \textit{network cycle duration} $T_{i}^{\text{nc}}$, and (iii) the size of the \textit{transmission window} $W_{i,\text{\gls{DC}}}$.

Let us define the \textit{uncertainty interval} $t_{\text{\gls{DC}}}^{\text{uni}}$ as the range of possible delays a packet from \gls{DC} flow may experience when traversing the \gls{5G}-\gls{TSN} network:
\begin{equation}
t_{\text{\gls{DC}}}^{\text{uni}} =\left[ \min(\widetilde{d}_{\text{\gls{DC}}}^{\text{emp}}),\ \hat{D}_{\text{\gls{DC}},p}^{\text{emp}}\right].
\label{eqn:uncertainty-interval}
\end{equation}
\indent This interval is bounded by the minimum and maximum values of the per-packet $\widetilde{d}_{\text{\gls{DC}}}^{\text{emp}}$. Note that we consider the $p\text{-th}$ percentile of the \gls{ZWSL} empirical delay distribution as the maximum value of the \textit{uncertainty interval}. Accordingly, we define the induced jitter of the \gls{5G}-\gls{TSN} network, $t_{\text{\gls{DC}}}^{\text{jit}}$, as the difference between the limits of the uncertainty interval:
\begin{equation}
t_{\text{\gls{DC}}}^{\text{jit}} = \max(t_{\text{\gls{DC}}}^{\text{uni}}) - \min(t_{\text{\gls{DC}}}^{\text{uni}}).
\label{eq:5Gjitter}
\end{equation}
\indent To guarantee a \textit{deterministic transmission}, two timing conditions must be satisfied between \gls{MS} and \gls{SL}:

\textit{First Condition for Determinism:} To ensure this one-to-one correspondence between the \textit{transmission windows} at \gls{MS} and \gls{SL}, the start time of the \textit{transmission window} at the \gls{SL} switch, i.e.,  $\delta'_{\text{\gls{DC}}}$, must satisfy the two boundary conditions valid for any \textit{network cycle} in Eq.~\eqref{eqn:cond_general1} or, alternatively, those in Eq.~\eqref{eqn:cond_general2}.
\begin{equation}
   \begin{split}
       &\text{C1: } \max\left(t_{\text{\gls{DC}}}^{\text{uni}}\right) \leq \delta'_{\text{\gls{DC}}}, \\
       &\text{C2: } \delta'_{\text{\gls{DC}}} + W_{i,\text{\gls{DC}}} \leq \min\left(t_{\text{\gls{DC}}}^{\text{uni}}\right) + T_i^{\text{nc}}.
   \end{split}
   \label{eqn:cond_general1}
\end{equation}
\indent The condition C1 indicates that the \textit{transmission window} at the \gls{SL} switch must start only after the last packet of the burst transmitted by the \gls{MS} switch has arrived in the current \textit{network cycle}, ensuring that all those packets are already available when the window opens. The condition C2 requires that the \textit{transmission window} must close before the first packet served in a subsequent \textit{network cycle} at the \gls{MS} switch arrives at the \gls{SL} switch in the next \textit{network cycle}.
\begin{equation}
   \begin{split}
       &\text{C3: } \max\left(t_{\text{\gls{DC}}}^{\text{uni}}\right) \leq \delta'_{\text{\gls{DC}}} + T_i^{\text{nc}}, \\
       &\text{C4: } \delta'_{\text{\gls{DC}}} + W_{i,\text{\gls{DC}}} \leq \min\left(t_{\text{\gls{DC}}}^{\text{uni}}\right).
   \end{split}
   \label{eqn:cond_general2}
\end{equation}
\indent The condition C3 stipulates that the \textit{transmission window} at the \gls{SL} switch can start only after the last packet of the burst transmitted by the \gls{MS} switch has arrived in the previous \textit{network cycle}. The condition C4 designates that the \textit{transmission window} must close before the first packet served in a subsequent \textit{network cycle} at the \gls{MS} switch arrives at the \gls{SL} switch in the current \textit{network cycle}. 
\color{black}

Either Eq.~\eqref{eqn:cond_general1} or Eq.~\eqref{eqn:cond_general2} ensures that all $N_{\text{\gls{DC}}}$ packets can be forwarded within a single \textit{transmission window}. Otherwise, the burst will necessarily be split across multiple \textit{transmission windows}, violating the determinism requirement. We call this effect \gls{ICI}, where the packets scheduled in a \textit{network cycle} may interfere with the ones scheduled in the next \textit{network cycle}.

\begin{figure*}[t!]
    \centering
   \includegraphics[width=0.972\textwidth]{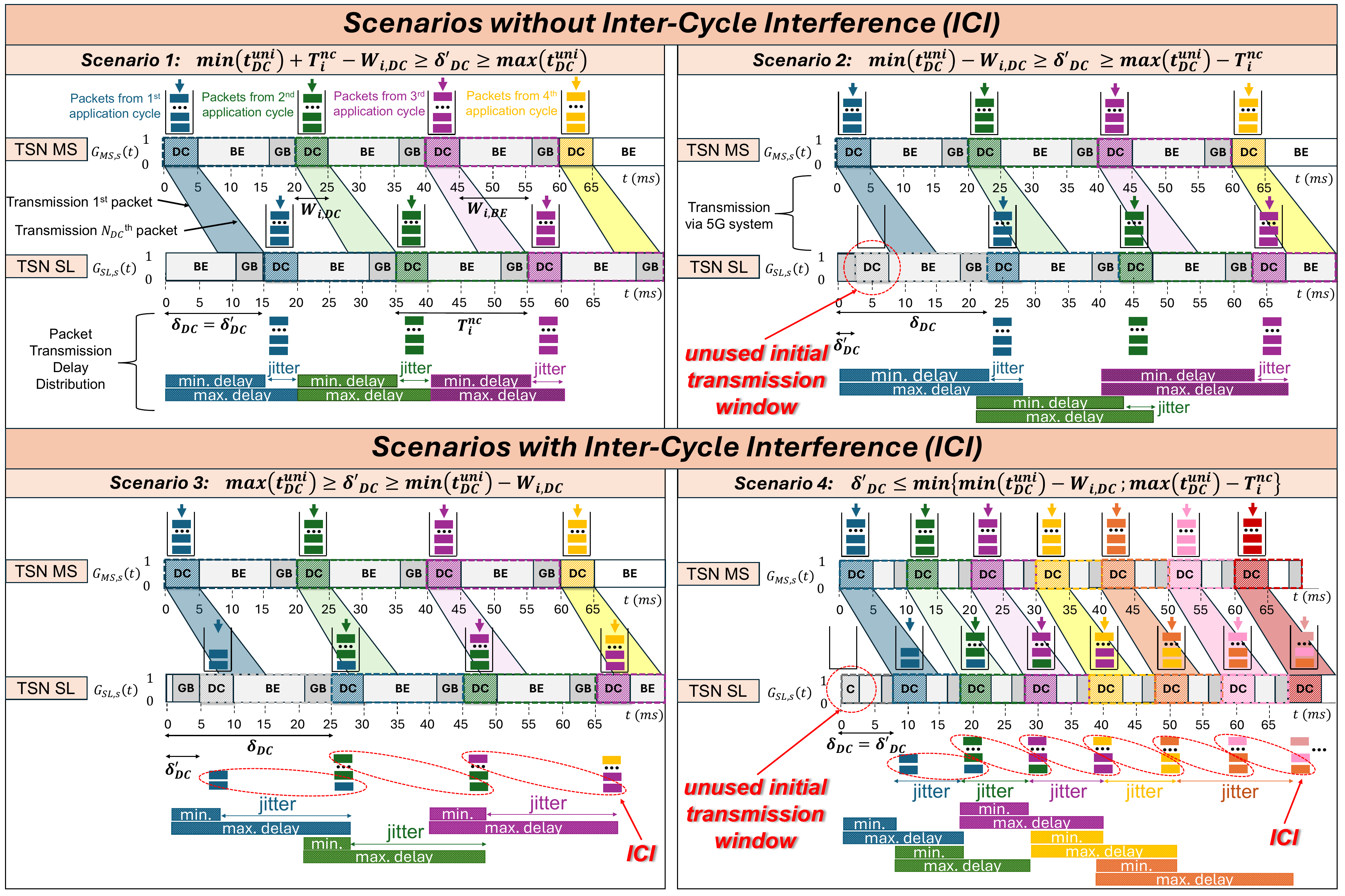}
    \caption{Impact of the \gls{5G} delay over \gls{TAS} cycles. \textbf{All Scenarios:} $\max\left(t_{\text{\gls{DC}}}^{\text{uni}}\right)=15$~ms, $\min\left(t_{\text{\gls{DC}}}^{\text{uni}}\right)=7.5$~ms and $W_{i,\text{\gls{DC}}}=5$~ms $\forall i \in \mathcal{I}^{\text{\gls{TSN}}}$. \textbf{Scenario~1:} $T_i^{\text{nc}}=20$~ms $\forall i \in \mathcal{I}^{\text{\gls{TSN}}}$ and $\delta_{\text{\gls{DC}}}=15$~ms. \textbf{Scenario~2:} $T_i^{\text{nc}}=20$~ms $\forall i \in \mathcal{I}^{\text{\gls{TSN}}}$ and $\delta_{\text{\gls{DC}}}=22.5$~ms. \textbf{Scenario~3:} $T_i^{\text{nc}}=20$~ ms $\forall i \in \mathcal{I}^{\text{\gls{TSN}}}$ and $\delta_{\text{\gls{DC}}}=25$~ms. \textbf{Scenario~4:} $T_i^{\text{nc}}=10$ ms $\forall i \in \mathcal{I}^{\text{\gls{TSN}}}$ and $\delta_{\text{\gls{DC}}}=7.5$~ ms.}
   \label{fig:5G_jitter_effect}
\end{figure*}

As $\max\left(t_{\text{\gls{DC}}}^{\text{uni}}\right) = \ \hat{D}_{\text{\gls{DC}},p}^{\text{emp}}$, in Eq.~\eqref{eqn:cond_general1} and Eq.~\eqref{eqn:cond_general2} the \textit{transmission window} forwarding all $N_{\text{\gls{DC}}}$ packets at the \gls{SL} switch is lower bounded by the $p\text{-th}$ percentile of the delay distribution of the \gls{ZWSL} empirical delay, $\widetilde{d}_{\text{\gls{DC}}}^{\text{emp}}$, and hence by the $p\text{-th}$ percentile of \gls{5G} system delay distribution. As a relevant consequence, the packet empirical delay $d_{\text{\gls{DC}}}^{\text{emp}}$ may increase in exchange for achieving \textit{deterministic transmission} according to the $p\text{-th}$ percentile and the \textit{network cycle} duration.

\textit{Second Condition for Determinism:} Directly comparing the upper and lower bounds for $\delta'_{\text{\gls{DC}}}$ isolated on one side of each of the inequalities, either in Eq.~\eqref{eqn:cond_general1} or Eq.~\eqref{eqn:cond_general2}, leads to the additional condition in Eq.~\eqref{eqn:necessary_condition}, which relates how the \gls{TAS} parameters must be configured with respect to the \gls{5G} jitter to guarantee \textit{deterministic transmission}.
\begin{equation}
    T_i^{\text{nc}} - W_{i,\text{\gls{DC}}} \geq t_{\text{\gls{DC}}}^{\text{jit}}.
    \label{eqn:necessary_condition}
\end{equation}

 Eq.~\eqref{eqn:necessary_condition} imposes a second fundamental condition on the interplay between the \gls{TAS} configuration and the statistical behavior of the \gls{5G} system.  It establishes that the \gls{5G}-induced jitter imposes a lower bound on the \textit{network cycle} $T_i^{\text{nc}}$. Additionally, increasing the \textit{transmission window} $W_{i,\text{\gls{DC}}}$ not only requires larger \textit{network cycle} $T_i^{\text{nc}}$ as Eq.~\eqref{eqn:necessary_condition} shows, but it will also indirectly increase $t_{\text{\gls{DC}}}^{\text{jit}}$ due to the cumulative queuing effect on packets in the \gls{5G} system. Furthermore, Eq.~\eqref{eqn:necessary_condition} also imposes a limit on the link utilization for \gls{DC} traffic, as additional \textit{transmission windows} of \gls{DC} traffic in the \textit{network cycle} period would cause \gls{ICI} and break the determinism.

The conditions derived above provide a foundation for \textit{deterministic transmission}. In the following subsection, we perform a detailed analysis of these conditions under different parameter configurations, identifying scenarios in which determinism is either achieved or violated to later be experimentally demonstrated in Section \ref{sec:Results}.

\subsection{Impact of \gls{5G} Delay on Offset Configuration}\label{sec:scenarios_cycle}
The analysis of the impact of the empirical delay $d_{\text{\gls{DC}}}^{\text{emp}}$, largely dominated by the \gls{5G} system, on the coordinated operation of \gls{TAS} mechanism in both the \gls{MS} and \gls{SL} switches is illustrated in Fig.~\ref{fig:5G_jitter_effect}, which shows the timing of data transmissions through the egress ports of the \gls{MS} and \gls{SL} switches interconnected via a \gls{5G} system. Each timeline is structured into consecutive \textit{network cycles}, each of which contains a single \textit{transmission window} allocated to the \gls{DC} flow. The figure considers four distinct scenarios based on the relative timing between the transmission windows at the \gls{SL} and the \textit{uncertainty interval} $t_{\text{\gls{DC}}}^{\text{uni}}$ in Eq. \eqref{eqn:uncertainty-interval}. These scenarios are defined by specific conditions on the network parameters $T_i^{\text{nc}}$, $W_{i,\text{\gls{DC}}}$, and the configured \textit{offset} $\delta_{\text{\gls{DC}}}$, which implicitly determines $\delta_{\text{\gls{DC}}}^{\prime}$ according to Eq.~\eqref{eqn:network_cycle_offset}. \\

\noindent \textbf{Scenario 1: Deterministic transmission with early arrival.}
    \begin{equation}
        \min(t_{\text{\gls{DC}}}^{\text{uni}}) + T_i^{\text{nc}} - W_{i,\text{\gls{DC}}} \geq \delta'_{\text{\gls{DC}}} \geq \max(t_{\text{\gls{DC}}}^{\text{uni}}).
        \label{eqn:scenario_1}
    \end{equation} 
From the previous conditions C1 and C2 in Eq.~\eqref{eqn:cond_general1}, the lower bound $\delta'_{\text{\gls{DC}}} \geq \max\left(t_{\text{\gls{DC}}}^{\text{uni}}\right)$, and the upper bound, $\delta'_{\text{\gls{DC}}} \leq \min\left(t_{\text{\gls{DC}}}^{\text{uni}}\right) + T_i^{\text{nc}} - W_{i,\text{\gls{DC}}}$, can be yielded for $\delta'_{\text{\gls{DC}}}$ in Eq.~\eqref{eqn:scenario_1}. With this, all packets from a given \textit{network cycle} arrive at the \gls{SL}
before the initial \textit{transmission window} opens. As a result, they can be transmitted entirely within that \textit{transmission window}.
This configuration ensures deterministic behavior in the \gls{5G}-\gls{TSN} network, with packet jitter bounded by the duration of the \textit{transmission window}, $W_{i,\text{\gls{DC}}}$. \\

\noindent \textbf{Scenario 2: Deterministic transmission with unused initial transmission window}. 
\begin{equation}
\begin{split}
    \min(t_{\text{\gls{DC}}}^{\text{uni}}) -W_{i,\text{\gls{DC}}} \geq \delta'_{\text{\gls{DC}}} \geq \max(t_{\text{\gls{DC}}}^{\text{uni}}) - T_i^{\text{nc}}.
\end{split}
\label{eqn:scenario_2}
\end{equation}
Now, from conditions C3 and C4 in Eq.~\eqref{eqn:cond_general2}, the lower bound $\delta'_{\text{\gls{DC}}} \geq \max\left(t_{\text{\gls{DC}}}^{\text{uni}}\right)-T_i^{\text{nc}}$, and the upper bound, $\delta'_{\text{\gls{DC}}} \leq \min\left(t_{\text{\gls{DC}}}^{\text{uni}}\right) - W_{i,\text{\gls{DC}}}$, can be yielded for $\delta'_{\text{\gls{DC}}}$ in Eq.~\eqref{eqn:scenario_2}.
No packets arrive before the initial \textit{transmission window} at \gls{SL} closes, which therefore remains unused. However, all packets are available before the second \textit{transmission window} opens, allowing their complete transmission. This results in higher minimum and maximum packet transmission delays compared to \textit{Scenario~1}, increased by the waiting in the queue until the next \textit{network cycle}. Nevertheless, the transmission remains deterministic, with bounded jitter. \\

\noindent \textbf{Scenario 3: Non-deterministic transmission with partial packet arrival.}
\begin{equation}
   \max(t_{\text{\gls{DC}}}^{\text{uni}}) \geq \delta'_{\text{\gls{DC}}} \geq \min(t_{\text{\gls{DC}}}^{\text{uni}}) - W_{i,\text{\gls{DC}}}.
\end{equation}
Some packets arrive in time to be transmitted during the initial \textit{transmission window} at the \gls{SL}, while others must wait for the second \textit{transmission window}. This results in \gls{ICI}, as defined in Section \ref{sec:determinism_condition}. Consequently, jitter increases to at least one full \textit{network cycle}, thereby affecting packets scheduled in the next \textit{network cycle}, and determinism is lost in the \gls{5G}-\gls{TSN} network. \\

\noindent \textbf{Scenario 4: Non-deterministic transmission with delayed arrival.} 
\begin{equation}
\begin{split}
    \delta'_{\text{\gls{DC}}} \leq
    \min\left\{ \min(t_{\text{\gls{DC}}}^{\text{uni}}) - W_{i,\text{\gls{DC}}},\right.  \left. \max(t_{\text{\gls{DC}}}^{\text{uni}}) - T_i^{\text{nc}} \right\}.
\end{split}
\end{equation}
This configuration represents the most adverse condition for the \gls{5G}-\gls{TSN} network because Eq.~\eqref{eqn:necessary_condition} is not met. This means \gls{ICI} is unavoidable. 
In this case, the second or subsequent \textit{transmission windows} at \gls{SL} may close before all packets have arrived, so some packets may be transmitted in the next \textit{network cycle}, leading to the highest delays and jitter among all scenarios.
We reflect the case where \gls{ICI} is extended to a third \textit{transmission window} due to the accumulation of packets at the \gls{SL}'s buffer between \textit{network cycles}.

\section{Testbed and Experimental Setup}\label{sec:Testbed}
In this section, we describe the implemented \gls{5G}-\gls{TSN} testbed and the considered experimental setup.

\subsection{Testbed Description}\label{sec:testbed_description}
\begin{figure*}[t!]
    \centering
   \includegraphics[width=\textwidth]{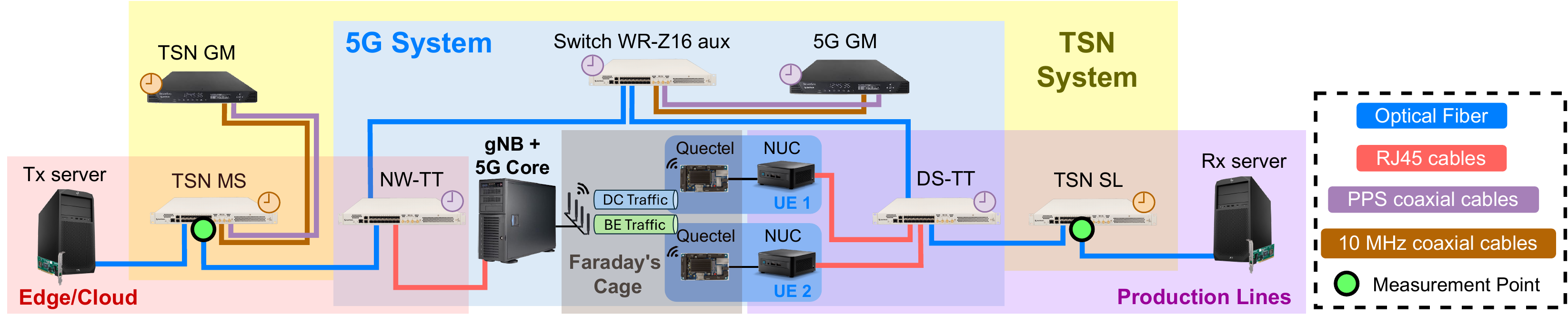}
    \caption{Proof of concept equipment and evaluated \gls{5G}-\gls{TSN} network scenario.}
   \label{fig:Testbed}
\end{figure*}

To carry out our empirical analysis, we implemented the testbed depicted in Fig.~\ref{fig:Testbed}. Its components are described below.

\textbf{\gls{5G} System.} The \gls{5G} network comprises a single \gls{gNB} and a \gls{5G} core, both implemented on a PC with a 50 MHz PCIe Amarisoft \gls{SDR} cards and an AMARI NW 600 license. The \gls{gNB} operates in the n78 band with 30 kHz subcarrier spacing and a bandwidth of 50 MHz. Data transmission uses a \gls{TDD} scheme with a pattern of four consecutive downlink slots, four uplink slots, and two flexible slots. Although our analysis focuses solely on downlink traffic, this configuration reserves resources for uplink, enabling a realistic testbed environment~\cite{AdamuzHinojosa2025}. Two \glspl{UE} are deployed, each consisting of a Quectel RM500Q-GL modem connected via USB to an Intel NUC 10 (i7-10710U, 16~GB RAM, 512 GB SSD) running Ubuntu 22.04. Experiments are conducted using one LABIFIX Faraday cage, with \gls{gNB} antennas connected to the \gls{SDR} via SMA connectors. Finally, although it is common to assign one \gls{DS-TT} per \gls{UE}~\cite{5GACIA-whitepaperI}, this proof of concept simplifies the setup by using a single \gls{DS-TT} for both \glspl{UE}. Similarly, we use a single \gls{NW-TT} for simplicity's sake.

\textbf{\gls{TSN} Network.} The \gls{TSN} network is built using Safran’s WR-Z16 switches. One switch operates as the \gls{MS}, another as the \gls{SL}, and two additional switches act as \gls{TSN} translators, i.e., \gls{NW-TT} and \gls{DS-TT}. The \gls{MS} is directly connected to a Safran SecureSync 2400 server, which provides the \gls{GM} clock to the \gls{SL} for time synchronization. Since the \gls{5G} system operates in \gls{PTP} \gls{TC} mode (implemented in \gls{TSN} translators \cite{caleya2025empirical}), an auxiliary WR-Z16 switch, also synchronized via a second SecureSync 2400, is used to distribute the \gls{5G} \gls{GM} clock between the \gls{TSN} translators. Each WR-Z16 switch is based on a Xilinx Zynq-7000 FPGA and a 1 GHz dual-core ARM Cortex-A9, enabling high switching rates and low processing delays under a Linux-based OS. The switches support IEEE 802.1Qbv \gls{TAS} and \glspl{VLAN}, and include sixteen 1GbE \gls{SFP} timing ports configurable as \gls{PTP} \gls{MS} or \gls{SL}. Each egress port provides four priority hardware queues to separate the different traffic flows, with a maximum buffer size of 6.6 kB per queue. This limits the number of \glspl{PCP} from 0 to 3, and also imposes a constraint on sustained throughput, as exceeding the draining capacity leads to packet drops. Additionally, timestamping probes on each port enables high-precision latency measurements between the output ports of the \gls{TSN} nodes.

\textbf{Testbed Clock Synchronization.} Time synchronization between the \gls{TSN} \gls{GM} clock server and the \gls{MS} is established via coaxial cables carrying two signals: a \gls{PPS} pulse for absolute phase alignment and a 10 MHz reference for frequency synchronization through oscillator disciplining. Similarly, the auxiliary WR-Z16 switch is synchronized with the \gls{5G} \gls{GM} clock server using the same coaxial interface, enabling accurate time distribution between the \gls{NW-TT} and \gls{DS-TT} to enable the \gls{TC} mode~\cite{caleya2025empirical}. In the testbed, the \gls{MS} and \gls{SL} communicate \gls{PTP} packets over IPv4 using unicast \gls{UDP} and the \gls{E2E} delay measurement mechanism. The \gls{PTP} transmission rate is configured to 1~packet per second.

\textbf{End Devices and Testbed Connections.} Two Ubuntu 22.04 LTS servers operate as packet generator with \textit{packETH} tool and sink, respectively. All components in the testbed are interconnected using 1 Gbps optical fiber links, except for the connections between the \gls{NW-TT}-\gls{gNB}, and \gls{DS-TT}-\glspl{UE}, which use 1 Gbps RJ-45 Ethernet cables.

\textbf{Network Traffic.} At the \gls{5G} core network, two distinct \glspl{DNN} are configured to create separate network slices for industrial traffic management. One carries both \gls{PTP} and \gls{DC} flows, while the other handles \gls{BE} flow, enabling differentiated routing and resource allocation. The \gls{5G} network employs IP transport because the considered \gls{UE} operates without Ethernet-based sessions. To support Layer 2 industrial automation traffic over IP, a \gls{VxLAN}-based tunneling mechanism is implemented~\cite{Ojcom_Oscar}, with two \glspl{VxLAN} configured accordingly: one transporting \gls{DC} and \gls{PTP} flows, and the other \gls{BE} flow. Packets are tagged with \gls{PCP} values reflecting the relative priority among the flows: \gls{PCP} 3 for packets of the \gls{PTP} flow, \gls{PCP} 2 for \gls{DC} flow packets, and \gls{PCP} 0 for \gls{BE} flow packets. Additionally, within the \gls{5G} network, \gls{5QI} values are assigned per flow’s packets, with 80 for \gls{PTP} and \gls{DC} traffic, and 9 for \gls{BE} traffic.

\subsection{Description of Experiments} \label{sec:experimental_scenarios}
We evaluate the packet transmission delay $d_{\text{\gls{DC}}}^{\text{emp}}$ for the \gls{DC} flow across five experimental scenarios. Each scenario analyzes a specific \gls{TAS} configuration parameter to evaluate its effect on the \gls{TSN} system's ability to tolerate \gls{5G}-induced delay. 

\textit{\textbf{Experiment 1:} Delay Analysis of \gls{5G} Network.} We analyze the effect of varying the traffic generation rate $R_{\text{\gls{DC}}}^{\text{gen}}$ on the delay and jitter of the \gls{5G} network to determine $\hat{D}_{\text{\gls{DC}},p}^{\text{emp}}$ and, with it, the \textit{uncertainty interval} $t_{\text{\gls{DC}}}^{\text{uni}}$. For that, we sweep $R_{\text{\gls{DC}}}^{\text{gen}}$ in 300 kbps increments from 350 kbps to 1.55 Mbps. For each $R_{\text{\gls{DC}}}^{\text{gen}}$, the \textit{transmission window} $W_{\text{\gls{MS}},\text{\gls{DC}}}$ is calculated based on the lower bound defined in Eq.~\eqref{eqn:window_bounds}, ensuring compliance with the WR-Z16's buffer size limitation. This results in \textit{transmission windows} at \gls{MS} ranging from 10.5~$\mu$s to 46.5~$\mu$s. \gls{TAS} is enabled at the \gls{MS}, while at \gls{SL} the output queue gate remains open 100\% of the time. This is done this way to estimate the \gls{ZWSL} empirical delay, $\widetilde{d}_{\text{\gls{DC}}}^{\text{emp}}$. The \textit{network cycle} is fixed at $T_{\text{\gls{MS}}}^{\text{nc}} = 30$~ms.

\textit{\textbf{Experiment 2:} Delay Analysis based on Offset between \textit{transmission windows} of \gls{MS} and \gls{SL} Switches}. We analyze the effect on $d_{\text{\gls{DC}}}^{\text{emp}}$ of different temporal shifts between \textit{network cycles} at \gls{MS} and \gls{SL}. \gls{TAS} is similarly configured at both switches, with fixed \textit{transmission window} $W_{i,\text{\gls{DC}}} = 46.5~\mu$s and \textit{network cycle} $T_{i}^{\text{nc}} = 30$~ms, $\forall i \in \mathcal{I}^{\text{\gls{TSN}}}$. We sweep \textit{offset} $\delta_{\text{\gls{DC}}}=\{5, 10, 15, 20, 25, 30\}$~ms.

\textit{\textbf{Experiment 3:} Delay Analysis Based on Network Cycle}. We study the influence of the \textit{network cycle} on $d_{\text{\gls{DC}}}^{\text{emp}}$ with a constant $\delta_{\text{\gls{DC}}}$ to analyze the scenarios described in Section \ref{sec:scenarios_cycle}. The \textit{network cycle} is varied in the range of $T_{i}^{\text{nc}} = \{6, 8, 10, 12.5, 15, 17.5, 20, 22.5\}$~ms
$\forall i \in \mathcal{I}^{\text{\gls{TSN}}}$. 
\textit{Transmission windows} are set to $W_{i,\text{\gls{DC}}}~=~\{9, 12, 15, 18, 22.5, 25.5, 30, 33\}$~$\mu$s,
 $\forall i \in \mathcal{I}^{\text{\gls{TSN}}}$, respectively, to keep the injected data rate into the \gls{5G}-\gls{TSN} network constant at 1.55 Mbps. 

\textit{\textbf{Experiment 4:} Delay Analysis considering Multiple Traffic flows with Same-Priority}. We evaluate the packet transmission delay when multiple distinct flows share the same priority output queue. Firstly, \gls{TAS} is enabled exclusively at the \gls{MS}, while at the \gls{SL}, the output queue gate remains open 100\% of the time, as in \textit{Experiment 1} to obtain $\widetilde{d}_{\text{\gls{DC}}}^{\text{emp}}$. The \textit{network cycle} is fixed at $T_i^{\text{nc}} = 30~\text{ms}$ $\forall i \in \mathcal{I}^{\text{\gls{TSN}}}$ and, to accommodate all the flows, \textit{transmission windows} are set to $W_{\text{\gls{MS}},\text{\gls{DC}}} = \{0.25, 0.5, 0.75, 1, 1.25, 1.5, 1.75\}$~ms, forwarding from 1 to 7 aggregated \gls{DC} flows at source each and analyzing the delay distribution for one of them. Then, we also configure \gls{TAS} at \gls{SL} so that $W_{\text{\gls{MS}},\text{\gls{DC}}} = W_{\text{\gls{SL}},\text{\gls{DC}}}$ to characterize $d_{\text{\gls{DC}}}^{\text{emp}}$. The \textit{offset} $\delta_{\text{\gls{DC}}}$ is constant according to previous experiments.

\textit{\textbf{Experiment 5:} Delay Analysis Based on \gls{BE} Traffic Load}. We sweep the \gls{BE} packet generation rates $R_{\text{\gls{BE}}}^{\text{gen}} = $ \{600, 650, 700, 750, 800, 850, 900, 950, 980\}~Mbps to analyze how the \gls{BE} load affects the \gls{DC} traffic $\widetilde{d}_{\text{\gls{DC}}}^{\text{emp}}$ distribution. The \textit{network cycle} is fixed to $T_i^{\text{nc}} = 30~\text{ms}$ $\forall i \in \mathcal{I}^{\text{\gls{TSN}}}$ and the \textit{transmission window} is set only at \gls{MS}, with $W_{\text{\gls{MS}},\text{\gls{DC}}} = $~46.5$\mu$s.

\begin{table*}[htbp]
\centering
\caption{Summary of Experimental Parameters for \gls{5G}-\gls{TSN} Network.}
\begin{tabular}{ccccccccc}
\hline
\textbf{Experiment} & \textbf{Parameter under analysis} & \textbf{Range / Value} & \textbf{$T_i^{\text{nc}}$} & \textbf{$W_{\text{\gls{MS},\gls{DC}}}$ (µs)} & \textbf{$\delta_{\text{\gls{DC}}}$} & \textbf{$L_{\text{\gls{DC}}}$ (B)} & \textbf{$R^{\text{gen}}_{\text{\gls{BE}}}$}\\
\hline
Exp. 1 & \gls{DC} generation rate $R^{\text{gen}}_{\text{\gls{DC}}}$ & 350–1,550 kbps (step 300 kbps) & 30 ms & 10.5–46.5 & – & 200 & 30 Mbps\\
Exp. 2 & Offset $\delta_{\text{\gls{DC}}} $ & $\{5,10,15,20,25,30\}$ ms & 30 ms & 46.5 & variable & 200 & 30 Mbps\\
Exp. 3 & Network cycle duration $T_i^{\text{nc}}$ & $\{6,8,10,12.5,15,17.5,20,22.5\} $ ms & variable & 9–33 & 20 ms & 200 & 30 Mbps\\
Exp. 4 & Number of load \gls{DC} flows & 1–7 flows & 30 ms & 250–1,750 & – / 20 ms & 100 & none\\
Exp. 5 & \gls{BE} load $R^{\text{gen}}_{\text{\gls{BE}}}$ & 600-980 Mbps & 30 ms & 46.5 & – & 200 & variable\\
\hline
\multicolumn{8}{l}{\textbf{Notes:} $T_{\text{GB}}=6.26~\mu$s, $W_{\text{\gls{PTP}}}=160$ ns, $r_{\varepsilon_{i,j}}=$ 1 Gbps,  $\forall \varepsilon_{i,j} \in \mathcal{E} \setminus \{\varepsilon_{\text{\gls{gNB},\gls{UE}}}\}$.}
\end{tabular}
\label{tab:exp_params}
\end{table*}

Note the $T_i^{\text{nc}}$ values, unlike the \textit{Cyclic-Synchronous} applications in \cite{5GACIA-whitepaperI}, have been adapted to the capabilities of our \gls{5G}-\gls{TSN} experimental setup and, with it, the flow constraints to potentially avoid \gls{ICI} at first and thus allow observable delay variation across experiments. The purpose of this work is not to replicate an exact industrial configuration but to analyze the interaction between \gls{5G} delay and jitter and \gls{TAS} under a synchronized \gls{5G}-\gls{TSN} network.

Additionally, each run of the experiments has been executed for 33 minutes, discarding the samples captured during the first 3 minutes to ensure stable synchronization between \gls{TSN} devices after clock locking. This time interval allows us to capture an average of 340,000 valid samples for a single \gls{DC} flow.

\subsection{Experimental Setup}\label{sec:experimental_setup}
In our experiments, the following configurations have been applied to the testbed.

\textbf{Traffic Generation and Configuration.} Focusing on each traffic flow type:

\begin{itemize} 
    \item \textit{\gls{DC} flow}: In \textit{Experiments 1-3} and \textit{5}, we use a single instance of \textit{packETH} to generate a \gls{DC} flow with packet size fixed at $L_{\text{\gls{DC}}}=200$~Bytes. Despite $T_i^{\text{nc}}$ being in the order of tens of milliseconds, \gls{DC} packets are generated every 750~$\mu$s to prevent the queue at \gls{MS} from emptying and therefore emulate a burst of packets within the same \textit{transmission window}. Then, $R_{\text{\gls{DC}}}^{\text{gen}} \propto W_{i,\text{\gls{DC}}}$. Our work focuses on \gls{TAS} configurations so that \gls{DC} flow has no particular application period, but is imposed by the opening of the queue at \gls{MS}, thus $T_{\text{\gls{DC}}}^{\text{app}}=T_{i}^{\text{nc}}$. 
    In \textit{Experiment~4}, we use multiple instances of \textit{packETH} to generate multiple \gls{DC} traffic flows, each with the same \gls{PCP} value but different destination addresses for the disaggregation at \gls{SL} to different output ports, measuring $d_{\text{\gls{DC}}}^{\text{emp}}$ for just the \textit{target} \gls{DC} flow. 
    In this experiment, the packet size has been reduced to 100 Bytes and the generation rate of the \textit{target} \gls{DC} flow’s packets is lessened to one packet every $100~\mu s$, while the \textit{background} \gls{DC} is set to \textit{packETH}'s maximum bitrate for interlacing. 
    
    \item \textit{\gls{BE} flow}: The packet size is fixed at $L_{\text{\gls{BE}}}=1500$~Bytes and generated at a constant rate of 30~Mbps for the \textit{Experiments 1-3}. \textit{Experiment 4} has no \gls{BE} traffic to avoid interference with \gls{DC} traffic while \textit{Experiment 5} sweeps this rate from 600~Mbps to 980~Mbps. 
\end{itemize}

\textbf{\gls{TAS} scheduling.} We consider a single \textit{transmission window} $W_{i,\text{\gls{DC}}}$ $\forall i \in \mathcal{I}^{\text{\gls{TSN}}}$ reserved for \gls{DC} traffic. The \textit{transmission window} $W_{i,\text{\gls{BE}}}$ for \gls{BE} traffic is obtained by subtracting the \gls{DC} \textit{transmission window} $W_{i,\text{\gls{DC}}}$, the fixed 6.26~$\mu$s guard band $T^{\text{GB}}$ that precedes it, and the 160 ns $W_{i,\text{\gls{PTP}}}$ reserved for a single \gls{PTP} message, from the total \textit{network cycle} duration, $T_{i}^{\text{nc}}$.

\textbf{Delay Measurement.} The empirical delay $d_{\text{\gls{DC}}}^{\text{emp}}$ and the \gls{ZWSL} empirical delay $\widetilde{d}_{\text{\gls{DC}}}^{\text{emp}}$ are measured at the output ports of the \gls{TSN} switches \gls{MS} and \gls{SL}, as shown in Fig.~\ref{fig:Testbed} (green dots). Packet transmission delay is measured using WR-Z16 timestamp probes placed at the output ports of the \gls{TSN} switches \gls{MS} and \gls{SL}. These probes extract the sequence number, which is embedded in the first 4~Bytes of the \gls{UDP} payload, and log the departure timestamp to CSV files. Per-packet latency is calculated by matching sequence numbers from both switches and computing the timestamp difference. The proposed configuration achieves negligible packet loss.

\textbf{Data Capture.} All experiments were run for at least 30~minutes as described in Section \ref{sec:experimental_scenarios}, generating a sufficient number of samples to ensure statistically valid results. All datasets and scripts are made publicly available to foster reproducibility\footnote{The repository is publicly accessible at this \href{https://github.com/paroma96/IOTJ--Impact_of_5G_Latency_and_Jitter_on_TAS_Scheduling_in_a_5G-TSN_Network.git}{link}.}.

A summary of these experiments and their configurations can be found in Table \ref{tab:exp_params}.

\section{Performance Results}\label{sec:Results}
In this section, we analyze the results of the performed experiments according to the equipment and the scenarios raised within the previous Section \ref{sec:Testbed}.

Prior to \gls{TAS}-based experiments, we conducted an empirical comparison of latency and jitter between a standalone \gls{TSN} network and its integration with \gls{5G} for a windowed \gls{DC} flow. The size of the \gls{DC} flow packets is 200 Bytes, and the \gls{TAS} configuration used in both scenarios is $W_{\text{\gls{MS},\gls{DC}}}~=~46.5$~$\mu$s and $T_{\text{\gls{MS}}}^{\text{nc}}=$~30~ms. While for \gls{TSN} we obtained that $\max\{\widetilde{d}_{\text{\gls{DC}}}^{\text{emp}}\}=~40.53$~$\mu$s and $t_{\text{\gls{DC}}}^{\text{jit}}=~29.54$~$\mu$s ($p=$1), in the \gls{5G}-\gls{TSN} setup both rose to $\max\{\widetilde{d}_{\text{\gls{DC}}}^{\text{emp}}\}=~18.41$~ms and $t_{\text{\gls{DC}}}^{\text{jit}}=~10.5$~ms ($p=$0.999). These results corroborate the observation in Section \ref{sec:DelayComponents} and make the characterization of $d_{\text{\gls{DC}}}^{\text{emp}}$ a key input to the wireless-aware \gls{TAS} scheduling.

\subsection{Experiment 1: Delay Analysis of \gls{5G} Network}\label{res:JitterTh}

\begin{figure}[b!]
    \centering
   \includegraphics[width=\columnwidth]{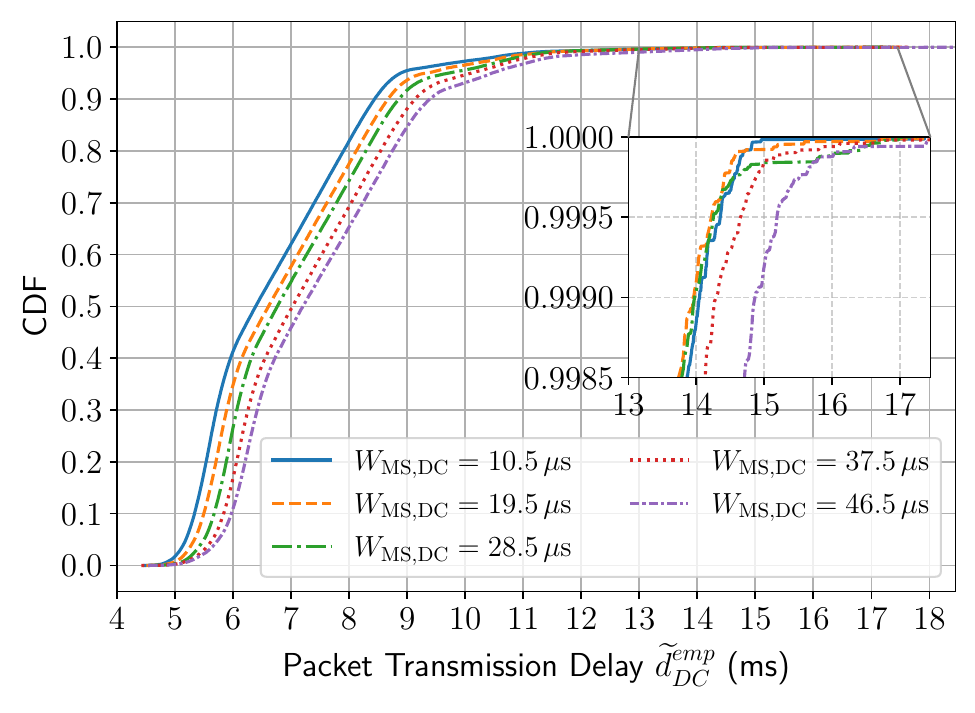}
    \caption{Empirical \gls{CDF} of $\widetilde{d}_{\text{\gls{DC}}}^{\text{emp}}$ for different \textit{transmission windows}  $W_{\text{\gls{MS}},\text{\gls{DC}}}$. \gls{TAS} is disabled in \gls{SL}.}
   \label{fig:5G_jitter_wdw}
\end{figure}

The resulting \glspl{CDF} of the \gls{ZWSL} empirical delay distribution, $F_{\widetilde{d}_{\text{\gls{DC}}}^{\text{emp}}}(\cdot)$, for the different \textit{transmission window} sizes $W_{\text{\gls{MS}},\text{\gls{DC}}}$ are presented in Fig.~\ref{fig:5G_jitter_wdw}. The results show that increasing $W_{\text{\gls{MS}},\text{\gls{DC}}}$ causes a moderate rightward shift in the \gls{CDF}, indicating higher $\widetilde{d}_{\text{\gls{DC}}}^{\text{emp}}$. This is because a larger \textit{transmission window} in the \gls{MS} allows more packets to be injected into the \gls{5G} system during each \textit{network cycle}. As more packets enter the \gls{5G} system, they accumulate in the buffer before transmission over the radio interface, leading to increased queuing delays and consequently higher $\widetilde{d}_{\text{\gls{DC}}}^{\text{emp}}$, as stated in Section \ref{sec:offset}.

For the evaluated \textit{transmission windows}, the distributions of $\widetilde{d}_{\text{\gls{DC}}}^{\text{emp}}$ show average delays between 6.39~ms and 7.21~ms, with a maximum observed delay of $\max \{t_{\text{\gls{DC}}}^{\text{uni}}\}~=~18.41$~ms. The 99.9th percentile is just below 15 ms, so we set the upper bound for the \gls{5G} delay contribution as $\hat{D}_{\text{\gls{DC}},0.999}^{\text{emp}}=~15~\text{ms}$. The observed minimum delay is $\min \{t_{\text{\gls{DC}}}^{\text{uni}}\} = 4.5$~ms. With this, the necessary condition for \textit{deterministic transmission} in Eq.~\eqref{eqn:necessary_condition} is satisfied: $T_{\text{\gls{MS}}}^{\text{nc}} - W_{\text{\gls{MS},\gls{DC}}} \approx 30$~ms $> t_{\text{\gls{DC}}}^{\text{jit}}=10.5$~ms. These results are aligned with the latency results in \cite{Damsgaard2023} and bounds are considered in subsequent experiments.

Despite these results, it is important to note that the obtained 99.9th percentile of the delay, $\hat{D}_{\text{\gls{DC}},0.999}^{\text{emp}}$, is not universal, as it depends on multiple factors such as the \gls{5G} configuration, the \gls{SINR}, the traffic load, etc. It must be estimated for any particular scenario and conditions where the \gls{5G} system is deployed. For example, the influence of the load is studied in the \textit{Experiments 4, 5}.

\subsection{Experiment 2: Delay Analysis Based on Offset between \textit{Transmission Windows} of \gls{MS} and \gls{SL} Switches}

\begin{figure}[t!]
    \centering
   \includegraphics[width=\columnwidth, height=4.38cm]{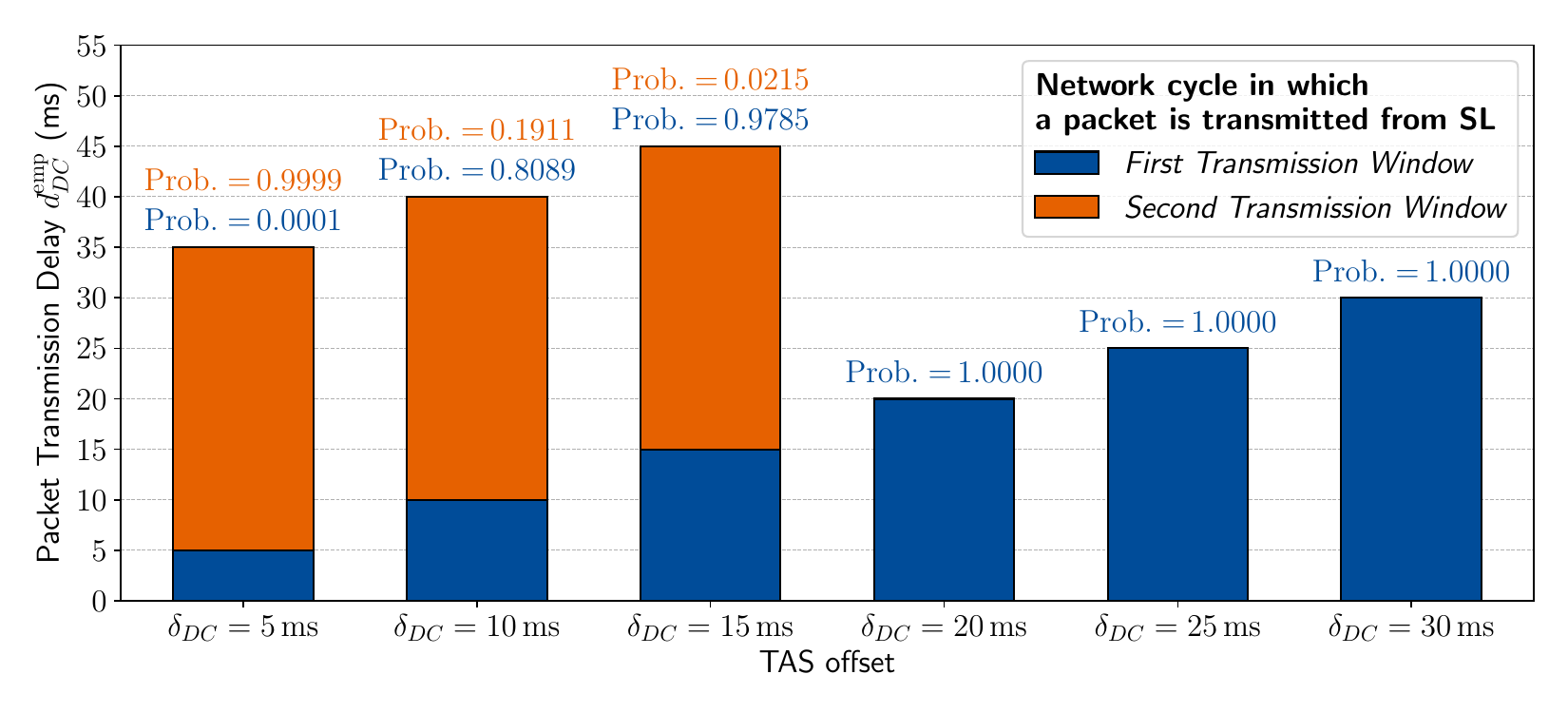}
    \caption{Empirical delay experienced by a packet arriving at the beginning of the $n$-th \textit{transmission window} at the \gls{SL} switch, considering different \textit{offset} $\delta_{\text{\gls{DC}}}$ values. Packets entering within the same \textit{transmission window} may experience individual theoretical delays in the range $\left[d_{\text{\gls{DC}}}^{\mathrm{emp}} - W_{i,\text{\gls{DC}}},\, d_{\text{\gls{DC}}}^{\mathrm{emp}} + W_{i,\text{\gls{DC}}}\right]$.}
   \label{fig:test_offset}
\end{figure}

Fig.~\ref{fig:test_offset} uses a grouped bar chart representation. Each evaluated scenario corresponds to a specific \textit{offset} $\delta_{\text{\gls{DC}}}$, and the plot represents the set of \textit{transmission windows} in the \gls{SL} switch as one or more bars, one per the $n$-th \textit{transmission window} used for transmitting an arbitrary packet in that scenario. The x-axis enumerates the evaluated scenarios, while the y-axis shows the minimum packet-transmission empirical delay $d_{\text{\gls{DC}}}^{\text{emp}}$ conditioned on the packet being transmitted in the $n$-th \textit{transmission window}. Furthermore, each bar is labeled with the probability that this case occurs. Note that the sum of the probabilities of all bars within the same evaluated scenario equals one, since they collectively cover all possible transmission outcomes for a specific \textit{offset} configuration.

Assuming that the necessary condition for achieving a \textit{deterministic transmission} in Eq.~\eqref{eqn:necessary_condition} is met, as seen in \textit{Experiment~1}, the next fundamental constraint to be satisfied is the boundary conditions in Eq.~\eqref{eqn:cond_general1} or, alternatively, in Eq.~\eqref{eqn:cond_general2}. With this, the configured \textit{offset} $\delta_{\text{\gls{DC}}}$ must be at least the 99th percentile of the \gls{ZWSL} empirical delay distribution that defines the upper bound of the \textit{uncertainty interval}, this is $\hat{D}_{\text{\gls{DC}},0.999}^{\text{emp}}=\max \{t_{\text{\gls{DC}}}^{\text{uni}}\}$. For \textit{network cycle offset} $\delta_{\text{\gls{DC}}}^{\prime} = \delta_{\text{\gls{DC}}} > \hat{D}_{\text{\gls{DC}},0.999}^{\text{emp}}$ (i.e., greater than 15 ms), 100\% of packets are transmitted within a single \textit{transmission window}, as evidenced by a single bar per case. These realizations correspond to the \textit{Scenario 1} depicted in Fig.~\ref{fig:5G_jitter_effect}. This indicates $d_{\text{\gls{DC}}}^{\text{emp}} \in [\delta_{\text{\gls{DC}}} - W_{i,\text{\gls{DC}}},\ \delta_{\text{\gls{DC}}} + W_{i,\text{\gls{DC}}}]$ $\forall i \in \mathcal{I}^{\text{\gls{TSN}}}$ since $\delta_{\text{\gls{DC}}}$ exceeds $\hat{D}_{\text{\gls{DC}},0.999}^{\text{emp}}$ and thus $\delta_{\text{\gls{DC}}}$ is statistically greater than the maximum delay of the \gls{5G} network, satisfying Eq. \eqref{eqn:offset_over_delay_distribution}. As $W_{i,\text{\gls{DC}}}$ is scaled accordingly (see Section \ref{sec:experimental_setup}), $d_{\text{\gls{DC}}}^{\text{emp}}=\delta_{\text{\gls{DC}}}$. Additionally, larger \textit{offsets} $\delta_{\text{\gls{DC}}}$ thus lead to higher latencies. 

When $ \delta_{\text{\gls{DC}}}^{\prime} = \delta_{\text{\gls{DC}}} \leq \hat{D}_{\text{\gls{DC}},0.999}^{\text{emp}}$ (i.e., equal or lower than 15 ms) not all packets arrive in time to be scheduled within the \textit{transmission window} in the same \textit{network cycle} at \gls{SL} and must therefore be deferred to the corresponding \textit{transmission window} of the next \textit{network cycle}. These realizations correspond to the \textit{Scenario 3} depicted in Fig.~\ref{fig:5G_jitter_effect}. The main consequence is that packets transmitted in the second \textit{transmission window} incur an additional delay approximately equal to $T_i^{\text{nc}}$. As a result, the empirical delay distribution becomes bimodal, meaning that a subset of packets are transmitted with a delay shifted by $T_i^{\text{nc}}$, i.e., $d_{\text{\gls{DC}}}^{\text{emp}} \in [\delta_{\text{\gls{DC}}} + T_i^{\text{nc}}  - W_{i,\text{\gls{DC}}}, \delta_{\text{\gls{DC}}} + T_i^{\text{nc}} + W_{i,\text{\gls{DC}}}]$. 

The setting of the 99.9th percentile \textit{offset} obtained from \textit{Experiment 1}, i.e., $\delta_{\text{\gls{DC}}}^{\prime}=\delta_{\text{\gls{DC}}}~=~{\hat{D}_{\text{\gls{DC}},0.999}^{\text{emp }}}=$~15~ms,  is not enough to transmit all packets within the same \textit{transmission window} due to the \gls{ICI} effect, increasing then the probability of being transmitted in a second \textit{network cycle}.

In conclusion, the \textit{offset} $\delta_{\text{\gls{DC}}}$ must be elected so that \gls{ICI} effect does not occur and, at the same time, it is not excessively large to increase latencies, i.e., $\delta_{\text{\gls{DC}}}=$~20~ms. However, as stated in Section \ref{sec:determinism_condition}, this higher \textit{offset} will unevitably increase the latency in exchange of guaranteeing the \textit{deterministic transmissions}.

\subsection{Experiment 3: Delay Analysis Based on Network Cycle}

\begin{figure}[t!]
    \centering
   \includegraphics[width=\columnwidth, height=4.6cm]{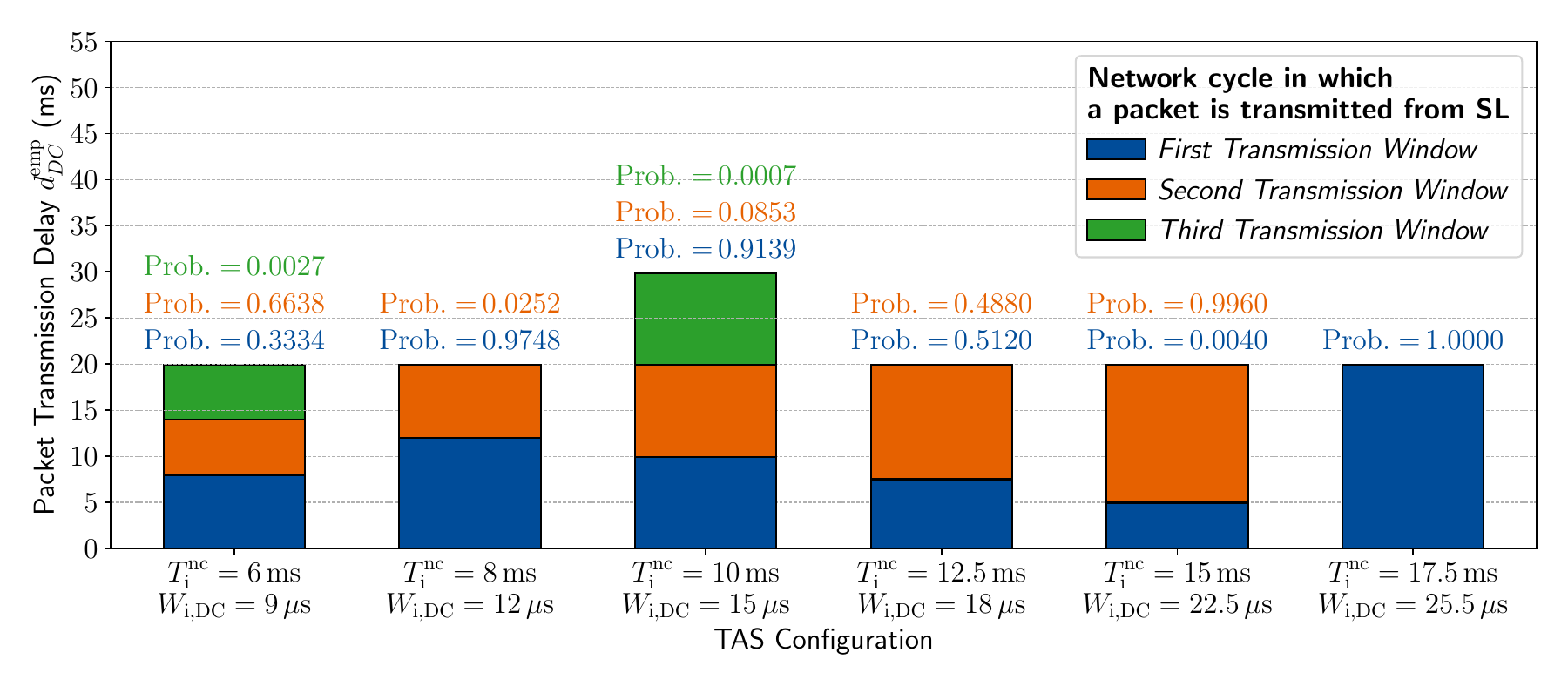}
    \caption{Empirical delay experienced by a packet arriving at the beginning of the $n$-th \textit{transmission window} at the \gls{SL} switch, considering different \gls{TAS} configurations.}
   \label{fig:test_2}
\end{figure}

Fig.~\ref{fig:test_2} uses the same grouped bar chart representation introduced in the previous experiment. Each evaluated scenario corresponds to a specific combination of the \textit{network cycle} $T_i^{\text{nc}}$ and the \textit{transmission window} size $W_{i,\text{\gls{DC}}}$. The represented realizations correspond to the \textit{Scenarios 2-4}, illustrated in Fig.~\ref{fig:5G_jitter_effect}, where $\delta_{\text{\gls{DC}}} = 20$~ms according to previous \textit{Experiment 2}. Some present severe \gls{ICI} as packets are transmitted from \gls{SL} across multiple \textit{transmission windows}. As the \textit{network cycle} $T_i^{\text{nc}}$ decreases, the percentage of packets transmitted in the target \textit{transmission window} also decreases. Consequently, the number of \textit{transmission windows} where packets of the same burst can be transmitted increases. For clarity, some evaluated \textit{network cycles} (i.e., $T_i^{\text{nc}} \geq 20$~ms, \textit{Scenario 1 in Fig.~\ref{fig:5G_jitter_effect}}) are not depicted in the Fig. \ref{fig:test_2} due to 100\% of generated packets being transmitted within a single \textit{transmission window}, i.e., with no \gls{ICI}, as seen in \textit{Experiment 2}.

On the one hand, most of the cases where $T_i^{\text{nc}}~<~\delta_{\text{\gls{DC}}}$ and $T_i^{\text{nc}}<\max\left\{t_{\text{\gls{DC}}}^{\text{uni}} \right\}$ $\forall i \in \mathcal{I}^{\text{\gls{TSN}}}$ may  see their transmissions split between \textit{network cycles}. This depends directly on $\delta_{\text{\gls{DC}}}^{\prime}$, as described in Eq. \eqref{eqn:network_cycle_offset}, e.g., $\delta_{\text{\gls{DC}}}^{\prime}=$\{7.5,~5\}~ms for $T_{i}^{\text{nc}}=\{12.5,~15\}$~ms, respectively, where $\delta_{\text{\gls{DC}}}^{\prime}~>~\min\left\{t_{\text{\gls{DC}}}^{\text{uni}} \right\}~-~W_{i,\text{\gls{DC}}}$ (\textit{Scenario~3} in Fig.~\ref{fig:5G_jitter_effect}). Nevertheless, the compliance with Eq. \eqref{eqn:necessary_condition} implies that a $\delta'_{\text{\gls{DC}}}$ correction may solve this \gls{ICI} and move on to \textit{Scenario~1}. For the cases where $T_{i}^{\text{nc}}=$\{6,~8,~10\}~ms, the condition of Eq.~\eqref{eqn:necessary_condition} is not met, i.e., $T_i^{\text{nc}}~-~W_{i,\text{\gls{DC}}}~<~t_{\text{\gls{DC}}}^{\text{jit}}=10.5$~ms. This means that \gls{ICI} effect is unavoidable. Moreover, given that $\delta_{\text{\gls{DC}}}^{\prime}=$\{2,~4,~0\}~ms, i.e., $\delta_{\text{\gls{DC}}}^{\prime}<\min\left\{t_{\text{\gls{DC}}}^{\text{uni}} \right\}-W_{i,\text{\gls{DC}}}$, these realizations fall under the \textit{Scenario 4} in Fig.~\ref{fig:5G_jitter_effect}. It results that minimum latency is equal to $\delta_{\text{\gls{DC}}}^{\prime}+T_{i}^{\text{nc}}$ as $\delta_{\text{\gls{DC}}}^{\prime}$ is not enough to accomplish the transmission of any packet within the initial \textit{transmission window}. 

On the other hand, when $T_i^{\text{nc}} < \delta_{\text{\gls{DC}}}$ and $T_{i}^{\text{nc}} > \max\left\{t_{\text{\gls{DC}}}^{\text{uni}} \right\}$  $\forall i \in \mathcal{I}^{\text{\gls{TSN}}}$, e.g., $T_i^{\text{nc}} = 17.5$~ms, an initial \textit{transmission window} opens at $\delta_{\text{\gls{DC}}}^{\prime} = 2.5$~ms, which is earlier than the \textit{transmission window} originally scheduled at $\delta_{\text{\gls{DC}}} = 20$~ms (\textit{Scenario 2} in Fig.~\ref{fig:5G_jitter_effect}). While these early \textit{transmission windows} may theoretically lead to \gls{ICI} if packets arrive prematurely, no such interference was observed. This is due to $\min\left\{t_{\text{\gls{DC}}}^{\text{uni}} \right\}~-~W_{i,\text{\gls{DC}}}~\geq~\delta_{\text{\gls{DC}}}^{\prime}$, preventing any packet from being transmitted from \gls{SL} before its planned \textit{transmission window}. As a result, 100\% of the packets are transmitted at $\delta_{\text{\gls{DC}}} = 20$~ms, consistent with the target scheduling. 

To conclude, shorter \textit{network cycles} $T_i^{\text{nc}}$ may lead to \gls{ICI} when the Eq. \eqref{eqn:necessary_condition} is not met. Furthermore, those packets queued at \gls{SL} before $\delta'_{\text{\gls{DC}}}$ suffer an empirical delay so that $d_{\text{\gls{DC}}}^{\text{emp}} < \delta_{\text{\gls{DC}}}$ and $d_{\text{\gls{DC}}}^{\text{emp}} > \delta_{\text{\gls{DC}}}$. This occurs when the \textit{network cycle offset} $\delta'_{\text{\gls{DC}}}$ is not enough, taking into account the $d_{\text{\gls{DC}}}^{\text{emp}}$ distribution, as stated in Section \ref{sec:determinism_condition}. In addition, this produces \gls{ICI} to the packets scheduled in the preceding and succeeding \textit{network cycles}, potentially preventing them from meeting their constraints. Then, $T_i^{\text{nc}} \geq  17.5$~ms. Nevertheless, considering a unique \gls{DC} flow type, $T_i^{\text{nc}}= T_{\text{\gls{DC}}}^{\text{app}}=30$~ms is kept for the \textit{Experiments 4-5}.

\subsection{Experiment 4: Delay Analysis Considering Multiple Traffic Flows with Same-Priority}

\begin{figure}[t!]
    \centering
    
    \begin{subfigure}[b]{\columnwidth}
        \centering
        \includegraphics[width=\columnwidth]{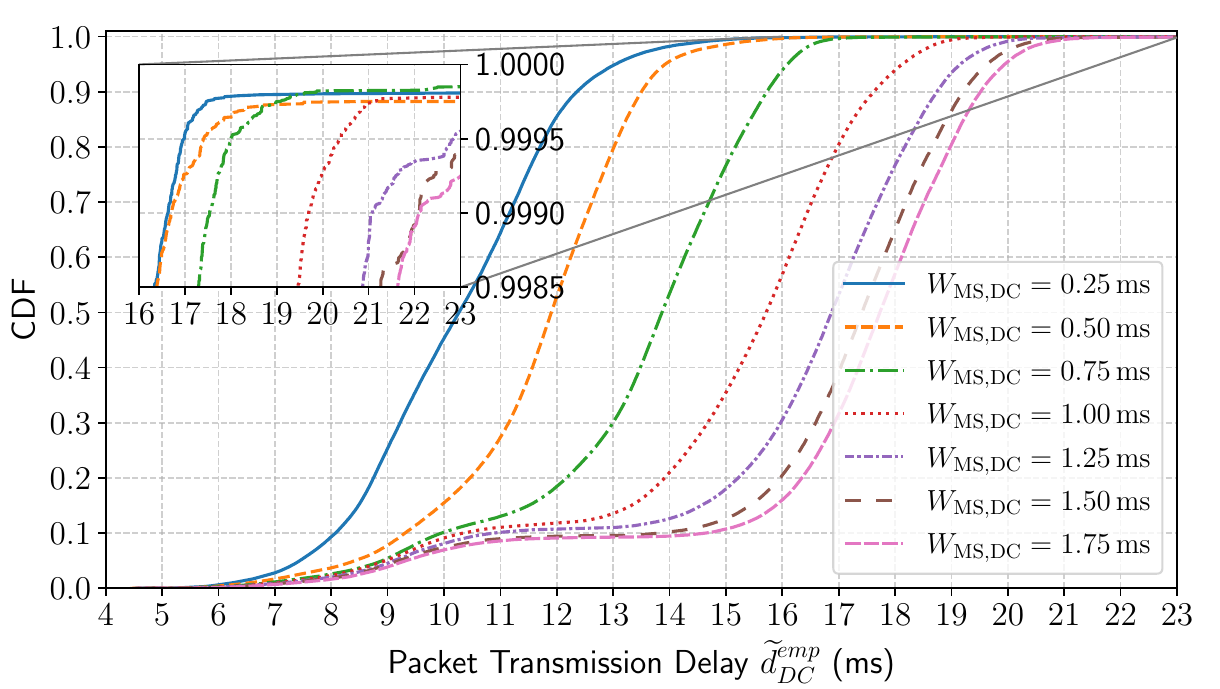}
        \caption{\gls{TAS} is disabled in \gls{SL}.}
        \label{fig:test_4a}
    \end{subfigure}

    \begin{subfigure}[b]{\columnwidth}
        \centering
        \includegraphics[width=\columnwidth]{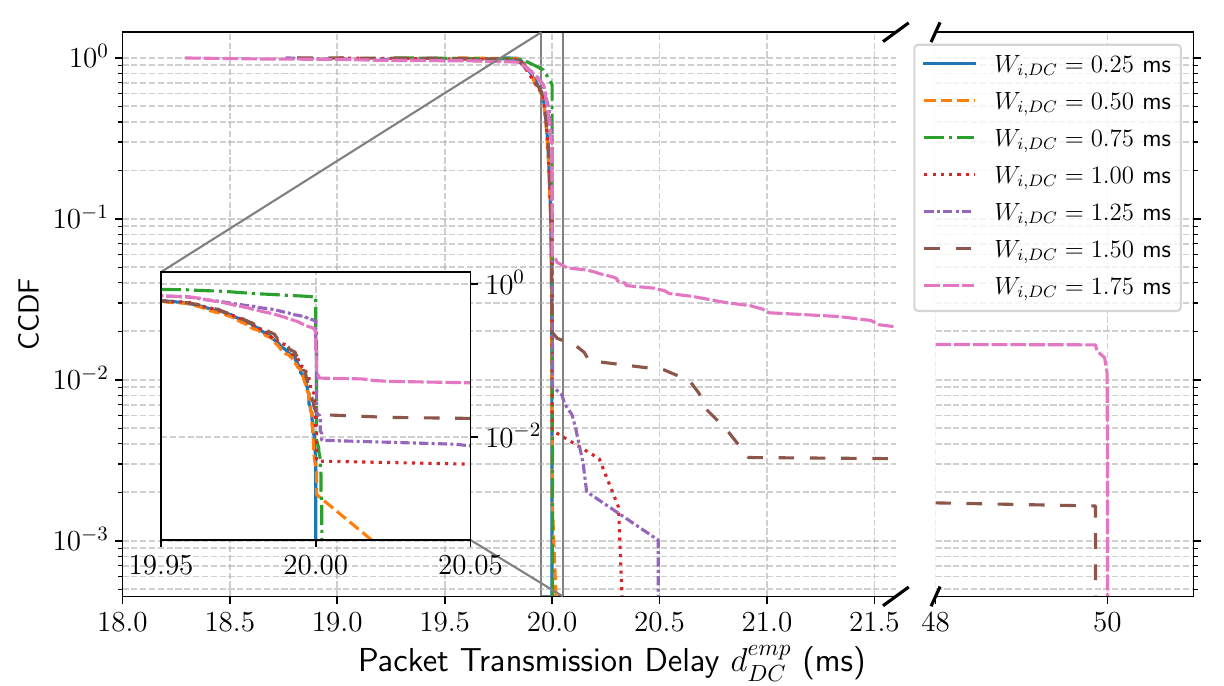}
        \caption{\gls{TAS} is enabled in \gls{SL}.}
        \label{fig:test_4b}
    \end{subfigure}
    
    \caption{Empirical \gls{CDF} of $\widetilde{d}_{\text{\gls{DC}}}^{\text{emp}}$ and \gls{CCDF} of $d_{\text{\gls{DC}}}^{\text{emp}}$ for different \textit{transmission windows}  $W_{\text{\gls{MS}},\text{\gls{DC}}}$.}
    \label{fig:test_subfiguras}
\end{figure}

In this experiment, \textit{target} and \textit{background} \gls{DC} flows share the same \textit{transmission window}. Two scenarios are carried out: one corresponding to the results shown in Fig.~\ref{fig:test_4a}, where \gls{TAS} is disabled at the \gls{SL}, measuring the \gls{5G} network delays for \textit{target} \gls{DC} traffic; while Fig.~\ref{fig:test_4b} illustrates the case where \gls{TAS} is enabled at \gls{SL}.

Similarly to the results presented in \textit{Experiment 1}, Fig.~\ref{fig:test_4a} shows the \glspl{CDF} of $\widetilde{d}_{\text{\gls{DC}}}^{\text{emp}}$ shift rightward as the duration of the \textit{transmission window} $W_{\text{\gls{MS}},\text{\gls{DC}}}$ increases. However, $W_{\text{\gls{MS}},\text{\gls{DC}}}$ is now significantly larger, reaching up to 1.75~ms, in contrast to the few tens of microseconds of \textit{Experiment 1}. As a consequence, significantly higher values of $\widetilde{d}_{\text{\gls{DC}}}^{\text{emp}}$ are observed. Although the minimum delay remains approximately $\min \{t_{\text{\gls{DC}}}^{\text{uni}}\} = 4.5$ ms, the average delays range from 10.27~ms to 17.79~ms. Additionally, the maximum observed value of $\widetilde{d}_{\text{\gls{DC}}}^{\text{emp}}$ exceeds 23~ms, while the 99.9th percentile in the worst-case configuration is $\hat{D}_{\text{\gls{DC}},0.999}^{\text{emp}} = 22~\text{ms}$. Consequently, $t_{\text{\gls{DC}}}^{\text{jit}}~=~17.5$~ms, which satisfies Eq.~\eqref{eqn:necessary_condition}. These results underline the increased \gls{5G} queuing delays and jitter induced by the presence of multiple concurrent \gls{DC} flows with the same priority in \gls{5G} downlink communications. 

Fig.~\ref{fig:test_4b} shows the \gls{CCDF} corresponding to the scenario where the \gls{TAS} mechanism is enabled at the \gls{SL}, where $d_{\text{\gls{DC}}}^{\text{emp}}$ is evaluated with $\delta_{\text{\gls{DC}}}=20$~ms, resulting from \textit{Experiment 2}. When $W_{i,\text{\gls{DC}}}~\in~\left[0.25, 0.75\right]$~ms $\forall~i~\in~\mathcal{I}^{\text{\gls{TSN}}}$, the measured delays are concentrated within the interval $[\delta_{\text{\gls{DC}}}-W_{i,\text{\gls{DC}}},~\delta_{\text{\gls{DC}}}]$ $\forall i \in~\mathcal{I}^{\text{\gls{TSN}}}$, for those packets that arrive in time to be scheduled within the \textit{transmission window} in the same \textit{network cycle} at the \gls{SL}. These latency values below $\delta_{\text{\gls{DC}}}$ happen when a packet at the \gls{MS} is transmitted at any time within the \textit{transmission window} $W_{i,\text{\gls{DC}}}$ $\forall i \in \mathcal{I}^{\text{\gls{TSN}}}$ and, due to packet disaggregation at the output ports in \gls{SL}, \textit{target} \gls{DC} packets waiting in the queue are quickly transmitted after $\delta_{\text{\gls{DC}}}$. Although some measures in $W_{i,\text{\gls{DC}}} \in \left[0.25, 0.75\right]$~ms $\forall i \in \mathcal{I}^{\text{\gls{TSN}}}$ are above $\delta_{\text{\gls{DC}}}=$ 20~ms, they cannot be attributed to any effect as they are within the amount allowed by the defined 99.9th percentile. Nevertheless, when $W_{i,\text{\gls{DC}}} \in \left[1, 1.25\right]$~ms $\forall i \in \mathcal{I}^{\text{\gls{TSN}}}$, the measured delays are concentrated within the interval $[\delta_{\text{\gls{DC}}}-W_{i,\text{\gls{DC}}},\delta_{\text{\gls{DC}}}+W_{i,\text{\gls{DC}}}]$ $\forall i \in \mathcal{I}^{\text{\gls{TSN}}}$. This occurs when the packets arrive later than those 20~ms but find their gate open during $W_{\text{\gls{SL}},\text{\gls{DC}}}$, and the probability of this effect increases as the $W_{i,\text{\gls{DC}}}$ $\forall i \in \mathcal{I}^{\text{\gls{TSN}}}$ gets higher. This means that $W_{\text{\gls{SL},\gls{DC}}} > N_{\text{\gls{DC}}} \cdot d_{\varepsilon_{\text{\gls{MS},\gls{NW-TT}}},\text{\gls{DC}} }^{\text{tran}}$ was necessary to guarantee the \textit{deterministic transmission} of certain \textit{target} \gls{DC} packets in exchange of reducing the bandwidth, although some jitter within $W_{i,\text{\gls{DC}}}$ spreads across subsequent \gls{TSN} nodes. Furthermore, this effect is highlighted in the cases of larger window sizes $W_{i,\text{\gls{DC}}} \in \left[1.5, 1.75\right]$~ms $\forall i \in \mathcal{I}^{\text{\gls{TSN}}}$—and then, greater aggregated traffic loads—, where packets start suffering greater latencies than $\delta_{\text{\gls{DC}}}+W_{i,\text{\gls{DC}}}$ $\forall i \in \mathcal{I}^{\text{\gls{TSN}}}$ and not all packets arrive in time to be transmitted within the same \textit{network cycle}. Consequently, some packets must be transmitted within the \textit{transmission window} of the following \textit{network cycle}, incurring an additional delay of $T_{i}^{\text{nc}} = 30$~ms, i.e., $\delta_{\text{\gls{DC}}}+T_{i}^{\text{nc}}=50$ ms. This behavior causes the \gls{ICI} effect.

In summary, in scenarios where multiple flows share the same priority, the solution for transmitting the packets of the \gls{DC} flows in a single \textit{transmission window} is to increase $W_{i,\text{\gls{DC}}}$ $\forall i \in \mathcal{I}^{\text{\gls{TSN}}}$ accordingly. However, this approach inevitably leads to increased jitter, which may become significant and impact the performance of the corresponding industrial application. Thus, in the \gls{SL}, the \textit{offset} $\delta_{\text{\gls{DC}}}$ should be set according to the new percentile $\hat{D}_{\text{\gls{DC}},0.999}^{\text{emp}}$ measured, as well as the \textit{transmission window} $W_{\text{\gls{SL}},\text{\gls{DC}}}$ should be resized to optimize bandwidth at the same time jitter is reduced.

\subsection{Experiment 5: Delay Analysis Based on \gls{BE} Traffic Load}

\begin{figure}[b]
    \centering
   \includegraphics[width=\columnwidth]{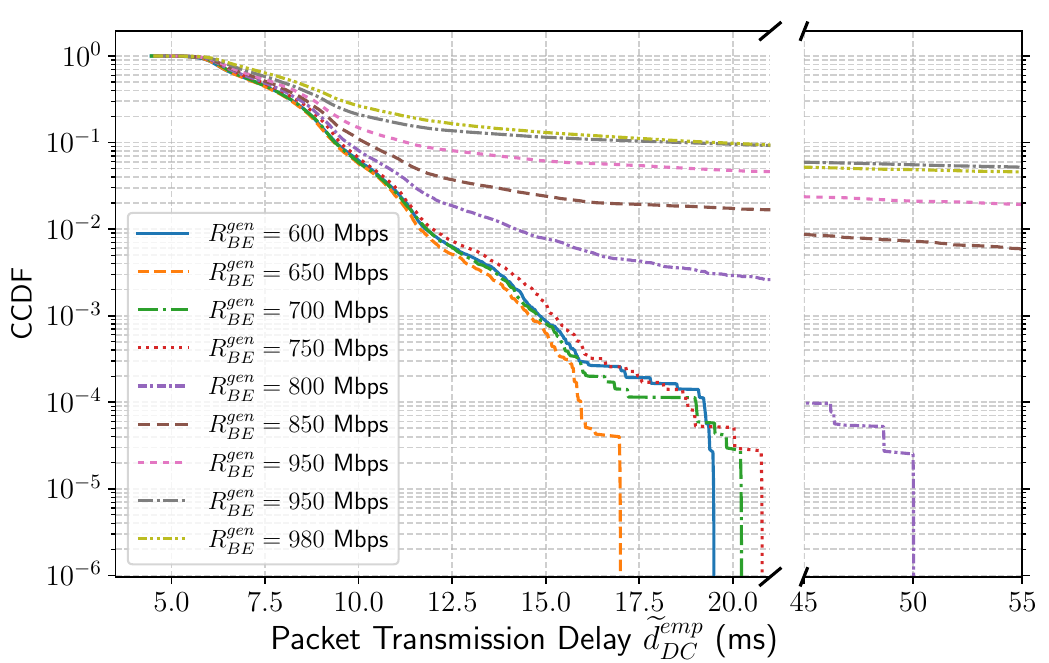}
    \caption{Empirical \gls{CCDF} of $\widetilde{d}_{\text{\gls{DC}}}^{\text{emp}}$ for different and higher \gls{BE} traffic generation rate $R^{\text{gen}}_{\text{\gls{BE}}}$.}
   \label{fig:res_exp5}
\end{figure}

The resulting \gls{CCDF} of $\widetilde{d}_{\text{\gls{DC}}}^{\text{emp}}$ is depicted in Fig. \ref{fig:res_exp5}, where a clear trend towards higher latencies can be seen as \gls{BE} load is increased. For the cases $R^{\text{gen}}_{\text{\gls{BE}}} \in [600, 650]$~Mbps, we obtain similar behavior of latencies as in the case of \textit{Experiment 1}, i.e., $\hat{D}_{\text{\gls{DC}},0.999}^{\text{emp}}\leq$~15~ms, so that we can also set  $\delta_{\text{\gls{DC}}}=$~20~ms, replicating \textit{Scenario 1} in Fig.~\ref{fig:5G_jitter_effect}. However, despite using the same \gls{TAS} configuration of \textit{Experiment~1} with fixed $W_{\text{\gls{MS}},\text{\gls{DC}}} = $~46.5$\mu$s, higher \gls{BE} loads such as $R^{\text{gen}}_{\text{\gls{BE}}} \in [700, 750]$~Mbps clearly triggers latencies slightly over $\delta_{\text{\gls{DC}}}$ such that few packets could not be transmitted until the next \textit{network cycle}. In those cases, the \textit{offset} should be rescaled up to, for example, $\delta_{\text{\gls{DC}}}=$~25~ms. Similarly, $R^{\text{gen}}_{\text{\gls{BE}}}=$~800~Mbps is enough for increasing latencies over 50~ms (\textit{Scenario 4} in Fig.~\ref{fig:5G_jitter_effect}). Additionally, latencies for $R^{\text{gen}}_{\text{\gls{BE}}} \in [850, 980]$~Mbps highly increase up to 800~ms, which is quite far from the industrial constraints. These results highlight the limited isolation between \gls{DC} and \gls{BE} traffic in the \gls{5G} system. Although $T^{\text{GB}}$ prevents collisions in the \gls{TAS} domain (Section~\ref{sec:TAS_model}), the \gls{5G} system only provides relative prioritization via the \gls{5QI} configuration. Consequently, resources are still shared, and under high \gls{BE} load, \gls{DC} packets may experience increased queuing delays due to buffer contention. Hence, the latency and jitter of the \gls{DC} flow are substantially increased by the \gls{BE} load, and the \textit{offset} $\delta_{\text{\gls{DC}}}$ must be reviewed again.

\section{Related Works}\label{sec:related_works}
This section reviews existing work on \gls{5G}-\gls{TSN} integration, with a specific focus on \gls{TAS} scheduling. In the literature, \gls{TSN} has been explored both as a fronthaul/backhaul solution within a \gls{5G} network and in scenarios where the \gls{5G} network acts as a \gls{TSN} bridge. Regarding the latter, we analyze works that address \gls{TAS}-based integration through architectural frameworks, simulation-based evaluations, and experimental testbeds. Finally, we compare our contributions with respect to the other works in each topic.

\subsection{Studies on \gls{TSN} with \gls{TAS} for \gls{5G} Fronthaul/Backhaul}
Some research efforts concentrate on the \gls{5G} fronthaul segment, which involves Ethernet-based low-latency transport solutions. Hisano et al.~\cite{Hisano2017} propose the gate-shrunk \gls{TAS}, a dynamic variant of \gls{TAS} that adjusts gate states via special control packets to enhance bandwidth efficiency without degrading delay for machine-to-machine communications. Nakayama et al.~\cite{Nakayama2021} develop an autonomous \gls{TAS} scheduling algorithm formulated as a boolean satisfiability problem that uses an FPGA-based solver for fast computation and flexible reconfiguration of the \gls{GCL} in response to changing traffic. Shibata et al.~\cite{Shibata2021} propose autonomous \gls{TAS} techniques, named iTAS and GS-TAS, and adaptive compression for mobile fronthaul to efficiently manage low-latency and bursty IoT traffic, achieving deterministic delay and supporting fronthaul and backhaul in \gls{5G} and IoT networks. 

Although these studies provide valuable \gls{TAS}-based solutions for deterministic low-latency fronthaul transport, they are not sufficient to ensure \gls{E2E} determinism in networks composed of both \gls{TSN} nodes and \gls{5G}, joined as a \gls{TSN} bridge.

\subsection{Surveys and Architectural Frameworks for \gls{TAS} in \gls{5G}-\gls{TSN} Networks}
From an architectural perspective, it is well established that the \gls{5G} system behaves as a \gls{TSN} logical switch, as discussed in \cite{Ros22}\cite{Lar20}. Several works address time synchronization \cite{Rod22} and \gls{5G}-\gls{TSN} QoS mapping \cite{Deb23} as key functions for this logical switch model. Comprehensive surveys and architectural frameworks have laid the foundation for understanding the role of \gls{TAS} in \gls{5G}-\gls{TSN} networks. Satka et al.~\cite{Sat23} provide an in-depth study that, while covering synchronization, delay, and security in \gls{5G}-\gls{TSN} systems, identifies \gls{TAS} as a critical yet underexplored component in achieving \gls{E2E} determinism. Egger et al.~\cite{Egger2025} highlight the incompatibilities between \gls{TAS}’s deterministic assumptions and the stochastic nature of wireless \gls{5G} links, advocating for a new “wireless-aware \gls{TSN} engineering” paradigm to adapt \gls{TAS} mechanisms for future \gls{5G} and \gls{6G} systems. Islam~\cite{Islam2024} applies graph neural networks combined with deep reinforcement learning for incremental joint \gls{TAS} and radio resource scheduling, illustrating the benefits of AI-driven optimization in complex integrated networks.  Nazari et al.~\cite{Nazari2023} develop the incremental joint scheduling and routing algorithm, emphasizing precise \gls{TAS} gate control and routing within centralized \gls{TSN} network configuration to minimize delay and packet delay variation.

While these contributions offer valuable architectural and conceptual perspectives on \gls{TAS} integration in \gls{5G}-\gls{TSN} networks, they lack empirical validation and do not specifically examine the impact of jitter on network performance.

\subsection{Simulation-Based Solutions for \gls{TAS} in \gls{5G}-\gls{TSN} Networks}
Several studies have relied on simulation to evaluate and improve \gls{TAS} scheduling, routing, and performance in \gls{5G}-\gls{TSN} integrated networks. Li et al.~\cite{Li2025} propose a fault-tolerant \gls{TAS} scheduling algorithm based on redundant scheduling and priority adjustment to reduce complexity and improve robustness against timing faults, offering a scalable baseline for \gls{5G}-\gls{TSN} integration. Wang et al.~\cite{Wang2024} propose the \textit{Balanced and Urgency First Scheduling (B-UFS)} heuristic algorithm to ensure deterministic \gls{E2E} delay for periodic time-critical flows. It introduces a pseudo-cyclic queuing and forwarding model for uncertain arrivals, a uniform resource metric, and a scheduling strategy that balances urgency and load across time and space to efficiently manage resources across the network. Debnath et al.~\cite{Deb23} present \textit{\gls{5G}TQ}, an open-source framework that enables \gls{5G}-\gls{TSN} integration through a \gls{TSN}-to-\gls{5G} \gls{QoS} mapping algorithm. It implements a \gls{QoS}-aware priority scheduler within the \gls{5G} MAC layer and evaluates \gls{RAN}-level scheduling strategies using ns-3, demonstrating improvements in delay and reliability for industrial traffic. Ginthör et al.~\cite{Ginthor2020} propose a constraint programming–based framework for optimizing \gls{E2E} flow scheduling in \gls{5G}-\gls{TSN} networks by modeling domain-specific constraints and a unified performance objective. Simulations on industrial topologies demonstrate improved schedulability and reduced delay compared to separate \gls{5G} and \gls{TSN} scheduling approaches. Chen et al.~\cite{Chen2022} explore the use of \gls{5G} as a \gls{TSN} bridge, integrating \gls{TAS} to support time-triggered flows across \gls{TSN} and \gls{5G} domains. It proposes a dynamic scheduling mechanism that allocates time slices to critical services, ensuring deterministic delay and jitter. However, the study abstracts away the wireless characteristics of \gls{5G}, focusing solely on its role as a deterministic forwarding bridge rather than analyzing radio-layer variability. Shih et al. \cite{Shih2023} propose a \gls{TAS} scheduling method based on constraint satisfaction that incorporates the variable residence time of the \gls{5G} logical bridge to preserve \gls{E2E} determinism. The approach models wireless timing uncertainty and introduces a robustness margin into the scheduling constraints to balance schedulability and reliability. Fontalvo-Hernández et al.~\cite{Fon24} analyze the feasibility of integrating \gls{5G} traffic into \gls{TSN} schedules governed by \gls{TAS}, focusing on jitter mitigation at the \gls{5G}-\gls{TSN} boundary. It evaluates the hold-and-forward buffering mechanism proposed in \gls{3GPP} standards, which equalizes packet residence time in \gls{5G} to make flows compatible with \gls{TAS} schedules. Using OMNeT++ simulations, the study quantifies the trade-off between jitter reduction and increased \gls{E2E} delay introduced by buffering. 

While these simulation-based studies provide valuable insights into \gls{TAS} scheduling and jitter mitigation, they lack practical guidelines for configuring \gls{TAS} to handle \gls{5G} delay variability, and do not validate their proposals in real environments. For instance, although the promising approach of the \textit{hold-and-forward} buffer jitter mitigation mechanism for \gls{TAS} presented by Fontalvo-Hernández et al. \cite{Fon24} and the complete \gls{5G}-\gls{TSN} architecture proposed by Debnath et al. \cite{Deb23}, both lack commercial \gls{5G} and \gls{TSN} equipment validation. Our work fills this gap by using a functional testbed to empirically analyze jitter impact and derive robust \gls{TAS} configurations for \gls{5G}-\gls{TSN} networks.

\subsection{Empirical Research on \gls{TAS} scheduling for \gls{5G}-\gls{TSN} networks}
Experimental validations complement theoretical and simulation results, providing practical insights into \gls{TAS} scheduling for \gls{5G}-\gls{TSN} networks. Jayabal et al.~\cite{Jayabal2025} design a contention-free \gls{CSMA} \gls{MAC} with transmission gating to minimize collisions and achieve low delay in \gls{5G}-\gls{TSN} scenarios. Agustí-Torra et al. \cite{torra24} aim to study architectural challenges and interoperability aspects in an emulated \gls{5G}-\gls{TSN} testbed. Aijaz et al.~\cite{Aij24} build a \gls{5G}-\gls{TSN} testbed using commercial \gls{TSN} and \gls{5G} devices to transmit traffic via IEEE 802.1Qbv \gls{TAS}. It evaluates \gls{E2E} delay and jitter by scheduling packets over a near product-grade \gls{5G} system under varying traffic and network conditions. The analysis offers a useful initial view of the impact of \gls{5G} integration on \gls{TAS} performance and outlines resource allocation strategies, though a more in-depth exploration remains open for future work. 

Recently, we investigated in~\cite{rodriguez2025} the impact of \gls{5G} network-induced delay and jitter on the performance of IEEE 802.1Qbv scheduling in integrated \gls{5G}-\gls{TSN} networks. This study involved an empirical analysis based on a real-world testbed, which included IEEE 802.1Qbv-enabled switches, \gls{TSN} translators, and a commercial \gls{5G} system. We focused on evaluating how the integration of \gls{5G} affects the deterministic behavior of IEEE 802.1Qbv scheduling, developed an experimental setup combining \gls{TSN} and \gls{5G} technologies, and identified key configuration parameters to optimize IEEE 802.1Qbv performance within a \gls{5G}-\gls{TSN} environment. However, neither in this nor in other empirical work are the conditions for deterministic communications defined, nor are the critical scenarios evaluated.

Although these works are characterized by also conducting an empirical testbed-based evaluation of a \gls{5G}-\gls{TSN} network with real traffic, they differ in scope and depth. Agustí-Torra et al. \cite{torra24} focus on preliminary design and implementation of the testbed without delving into the evaluation of the feasibility of combining for different \gls{TAS} configurations. In the case of Aijaz et al. \cite{Aij24}, the enwindowed traffic from a single \gls{TSN} switch is examined through the \gls{5G} system, focusing solely on delay performance rather than assessing a full \gls{TAS} configuration. Similarly, Jayabal et al. \cite{Jayabal2025} aim to enhance wireless \gls{TSN} \gls{MAC} coordination without integrating or characterizing real \gls{5G} latency behavior. In contrast, our work focuses on characterizing this delay for a specific \gls{TAS} configuration in order to determine the required \textit{offset} between \gls{TSN} nodes in the downlink for deterministic operation.

\section{Conclusions and Future Work}\label{sec:Conclusions}
In this work, we have characterized how \gls{5G}-induced delay and jitter affect the coordinated operation of IEEE 802.1Qbv \gls{TAS} in integrated \gls{5G}-\gls{TSN} networks, with the objective of determining the timing conditions required to preserve deterministic transmission. We consider deterministic transmission as the scenario in which all packets of the same application cycle are forwarded within a single transmission window at both \gls{TSN} switches adjacent to the \gls{5G} segment, ensuring bounded jitter not exceeding the transmission window.

To enable deterministic transmission, a temporal offset must be introduced between the network cycles of the \gls{TSN} switches enclosing the \gls{5G} segment, dimensioned from a high-percentile bound of the \gls{5G} empirical delay. In addition, four timing constraints must be satisfied to ensure that all packets from the same application cycle are confined within a single transmission window. We also revealed another fundamental condition: the difference between the network cycle duration and the configured transmission window must be strictly larger than the \gls{5G} jitter, establishing how \gls{TAS} parameters must be configured with respect to \gls{5G} delay to avoid \gls{ICI}. These conditions were validated using a commercial \gls{5G}-\gls{TSN} testbed under realistic equipment-induced delay variability. Furthermore, our experiments showed that multiple delay-critical flows sharing the same priority increase \gls{5G} queuing delays and jitter, requiring larger offsets and transmission windows to maintain application cycle confinement. The presence of best effort traffic further broadens the \gls{5G} delay distribution, even with \gls{TAS} correctly configured, demonstrating that flow concurrency, traffic load, and \gls{5G} queuing dynamics must be explicitly considered to preserve the deterministic transmission.

Our results call for investigating jitter-mitigation techniques, such as the hold-and-forward buffering mechanism, to alleviate the \gls{ICI} effect and achieve near-full link utilization under realistic \gls{5G} delay variability. In addition, the performance of the \gls{5G}-\gls{TSN} network can be further enhanced by leveraging \gls{uRLLC}-oriented latency reduction features such as configured grants, mini-slot scheduling, \gls{5G} network slicing, or upcoming 6G systems. Our findings also motivate the design of adaptive algorithms capable of dynamically adjusting the offset and \gls{TAS} parameters based on real-time traffic load, empirical \gls{5G} delay distribution, and radio channel conditions. Finally, future work could analyze how to guarantee determinism under isochronous traffic.

\bibliographystyle{ieeetr}
\bibliography{references}

\begin{thebibliography}{10}

\bibitem{Groshev2021}
M.~Groshev, C.~Guimarães, J.~Martín-Pérez, and A.~de~la Oliva, ``{Toward Intelligent Cyber-Physical Systems: Digital Twin Meets Artificial Intelligence},'' {\em IEEE Commun. Mag.}, vol.~59, no.~8, pp.~14--20, 2021.

\bibitem{Saad20}
W.~Saad, M.~Bennis, and M.~Chen, ``{A Vision of 6G Wireless Systems: Applications, Trends, Technologies, and Open Research Problems},'' {\em IEEE Netw.}, vol.~34, no.~3, pp.~134--142, 2020.

\bibitem{ieee8021q}
{IEEE 802.1 Working Group}, ``{IEEE Standard for Local and Metropolitan Area Network--Bridges and Bridged Networks},'' {\em IEEE, Standard Std}, vol.~802, 2018.

\bibitem{ieee8021qbv}
{IEEE 802.1 Working Group}, ``{IEEE Standard for Local and Metropolitan Area Networks -- Bridges and Bridged Networks - Amendment 25: Enhancements for Scheduled Traffic},'' {\em {IEEE Std 802.1Qbv-2015 (Amendment to IEEE Std 802.1Q-2014 as amended by IEEE Std 802.1Qca-2015, IEEE Std 802.1Qcd-2015, and IEEE Std 802.1Q-2014/Cor 1-2015)}}, pp.~1--57, 2016.

\bibitem{5GACIA-whitepaperI}
{5G-ACIA}, ``{Integration of 5G with Time-Sensitive Networking for Industrial Communications}.'' White Paper, Feb. 2021.

\bibitem{rodriguez2025}
P.~Rodriguez-Martin, O.~Adamuz-Hinojosa, P.~Muñoz, J.~Caleya-Sanchez, J.~Navarro-Ortiz, and P.~Ameigeiras, ``{Empirical Analysis of the Impact of 5G Jitter on Time-Aware Shaper Scheduling in a 5G-TSN Network},'' in {\em IEEE WFCS}, pp.~1--8, Jun 2025.

\bibitem{Darroudi2024}
S.~M. Darroudi, N.~Domènech, M.~Grandy, and R.~Guerra-Gomez, ``{On the TSN and 5G network integration approaches, 5G features proof, advantages and challenges},'' in {\em EuCNC/6G Summit}, pp.~688--693, 2024.

\bibitem{3gpp_ts_23_501_v19_0_0}
{3GPP TS 23.501}, ``{{System architecture for the 5G System (5GS) (Release 19)}},'' June 2024.

\bibitem{Ojcom_Oscar}
O.~Adamuz-Hinojosa, F.~Delgado-Ferro, J.~Navarro-Ortiz, P.~Muñoz, and P.~Ameigeiras, ``{Unleashing 5G Seamless Integration With TSN for Industry 5.0: Frame Forwarding and QoS Treatment},'' {\em IEEE Open J. Commun. Soc.}, vol.~6, pp.~4874--4884, 2025.

\bibitem{IIC_traffic}
{Industrial Internet Consortium and OpenFog}, ``{{Time Sensitive Networks for Flexible Manufacturing Testbed Characterization and Mapping of Converged Traffic Types}},'' white paper, {Industrial Internet Consortium}, Mar. 2019.
\newblock [Online]. Available: \url{https://www.iiconsortium.org/pdf/IIC_TSN_Testbed_Char_Mapping_of_Converged_Traffic_Types_Whitepaper_20180328.pdf}.

\bibitem{Damsgaard2023}
S.~B. Damsgaard, D.~Segura, M.~F. Andersen, S.~Aaberg~Markussen, S.~Barbera, I.~Rodríguez, and P.~Mogensen, ``{Commercial 5G NPN and PN Deployment Options for Industrial Manufacturing: An Empirical Study of Performance and Complexity Tradeoffs},'' in {\em IEEE PIMRC}, pp.~1--7, 2023.

\bibitem{Perdigao2024}
A.~Perdigão, J.~Quevedo, and R.~L. Aguiar, ``{Is Release 15 Ready for the Industry?},'' {\em IEEE Access}, vol.~12, pp.~17651--17668, 2024.

\bibitem{Oge2020}
Y.~Oge {\em et~al.}, ``{{Software-Based Time-Aware Shaper for Time-Sensitive Networks}},'' {\em {IEICE Trans. Commun.}}, vol.~103, no.~3, pp.~167--180, 2020.

\bibitem{Oliver2018}
R.~Serna~Oliver, S.~S. Craciunas, and W.~Steiner, ``{IEEE 802.1Qbv Gate Control List Synthesis Using Array Theory Encoding},'' in {\em IEEE RTAS}, pp.~13--24, 2018.

\bibitem{Walrand2023}
J.~Walrand, ``{A Concise Tutorial on Traffic Shaping and Scheduling in Time-Sensitive Networks},'' {\em IEEE Commun. Surv. Tutor.}, vol.~25, no.~3, pp.~1941--1953, 2023.

\bibitem{Lin22}
J.~Lin {\em et~al.}, ``{Rethinking the Use of Network Cycle in Time-Sensitive Networking (TSN) Flow Scheduling},'' in {\em IEEE/ACM IWQoS}, pp.~1--11, IEEE, 2022.

\bibitem{ieee8021qas}
{IEEE 802.1 Working Group}, ``{IEEE Standard for Local and Metropolitan Area Networks--Timing and Synchronization for Time-Sensitive Applications},'' {\em IEEE, Standard Std}, vol.~802, pp.~1--421, 2020.

\bibitem{StandadsIEEE1588_2019}
``{IEEE Standard for a Precision Clock Synchronization Protocol for Networked Measurement and Control Systems},'' {\em {IEEE Std 1588-2019}}, pp.~1--499.

\bibitem{striffler2021}
T.~Striffler and H.~D. Schotten, ``{The 5G Transparent Clock: Synchronization Errors in Integrated 5G-TSN Industrial Networks},'' in {\em IEEE INDIN}, pp.~1--6, 2021.

\bibitem{caleya2025empirical}
J.~Caleya-Sanchez {\em et~al.}, ``{Empirical Evaluation of a 5G Transparent Clock for Time Synchronization in a TSN-5G Network},'' in {\em IEEE PIMRC}, Sep. 2025.

\bibitem{munoz2025}
P.~Muñoz, P.~Rodriguez-Martin, J.~Caleya-Sanchez, J.~Prados-Garzon, O.~Adamuz-Hinojosa, and P.~Ameigeiras, ``{Joint Scheduling of IEEE 802.1AS gPTP and Industrial Data Traffic in TSN-6G Networks},'' {\em IEEE Access}, vol.~13, pp.~99842--99862, 2025.

\bibitem{3gpp_ts22104}
{3GPP Technical specification (TS) TS22.104}, ``{Service requirements for cyber-physical control applications in vertical domains (V17.7.0 Release 17)},'' 2022.

\bibitem{belliardi2018use}
R.~Belliardi, J.~Dorr, and T.~Enzinger, ``{Use cases IEC/IEEE 60802},'' {\em V1}, vol.~3, pp.~1--74, 2018.

\bibitem{Xue2024}
C.~Xue, T.~Zhang, Y.~Zhou, M.~Nixon, A.~Loveless, and S.~Han, ``{Real-Time Scheduling for 802.1Qbv Time-Sensitive Networking (TSN): A Systematic Review and Experimental Study},'' in {\em IEEE RTAS}, pp.~108--121, 2024.

\bibitem{Egger2025}
S.~Egger {\em et~al.}, ``{Wireless-Aware TSN Engineering: Implications for 5G and Upcoming 6G Networks},'' {\em IEEE Netw.}, vol.~39, no.~3, pp.~99--107, 2025.

\bibitem{AdamuzHinojosa2025}
O.~Adamuz-Hinojosa {\em et~al.}, ``{Empirical Analysis of 5G TDD Patterns Configurations for Industrial Automation Traffic},'' in {\em EuCNC/6G Summit}, pp.~1--6, June 2025.

\bibitem{Hisano2017}
D.~Hisano {\em et~al.}, ``{Gate-Shrunk Time Aware Shaper: Low-Latency Converged Network for 5G Fronthaul and M2M Services},'' in {\em IEEE GLOBECOM}, pp.~1--6, 2017.

\bibitem{Nakayama2021}
Y.~Nakayama, R.~Yaegashi, A.~H.~N. Nguyen, and Y.~Hara-Azumi, ``{Real-Time Reconfiguration of Time-Aware Shaper for ULL Transmission in Dynamic Conditions},'' {\em IEEE Access}, vol.~9, pp.~115246--115255, 2021.

\bibitem{Shibata2021}
N.~Shibata {\em et~al.}, ``{Time Sensitive Networking for 5G NR Fronthauls and Massive Iot Traffic},'' {\em J. Lightwave Technol.}, vol.~39, pp.~5336--5343, Aug 2021.

\bibitem{Ros22}
P.~M. Rost and T.~Kolding, ``{Performance of Integrated 3GPP 5G and IEEE TSN Networks},'' {\em IEEE Commun. Stand. Mag.}, vol.~6, no.~2, pp.~51--56, 2022.

\bibitem{Lar20}
A.~Larrañaga, M.~C. Lucas-Estañ, I.~Martinez, I.~Val, and J.~Gozalvez, ``{Analysis of 5G-TSN Integration to Support Industry 4.0},'' in {\em IEEE ETFA}, vol.~1, pp.~1111--1114, 2020.

\bibitem{Rod22}
S.~Rodrigues and J.~Lv, ``{Synchronization in Time-Sensitive Networking: An Introduction to IEEE Std 802.1AS},'' {\em IEEE Commun. Stand. Mag.}, vol.~6, no.~4, pp.~14--20, 2022.

\bibitem{Deb23}
R.~Debnath, M.~S. Akinci, D.~Ajith, and S.~Steinhorst, ``{5GTQ: QoS-Aware 5G-TSN Simulation Framework},'' in {\em IEEE VTC-Fall}, pp.~1--7, 2023.

\bibitem{Sat23}
Z.~Satka, M.~Ashjaei, H.~Fotouhi, M.~Daneshtalab, M.~Sjödin, and S.~Mubeen, ``{A comprehensive systematic review of integration of time sensitive networking and 5G communication},'' {\em J. Syst. Archit.}, vol.~138, p.~102852, 2023.

\bibitem{Islam2024}
S.~T. Islam, ``{WIP: AI-Based Dynamic Joint Schedule Calculation for TSN over 5G using GCN-TD3},'' in {\em IEEE ETFA}, pp.~1--4, 2024.

\bibitem{Nazari2023}
H.~K. Nazari, M.~A. Kurt, H.-H. Liu, S.~Senk, G.~T. Nguyen, and F.~H.~P. Fitzek, ``{Incremental Joint Scheduling and Routing for 5G-TSN Integration},'' in {\em Eur. Wirel. Conf}, pp.~110--116, 2023.

\bibitem{Li2025}
G.~Li, S.~Wang, Y.~Huang, T.~Huang, Y.~Cui, and Z.~Xiong, ``{Optimizing Fault-Tolerant Time-Aware Flow Scheduling in TSN-5G Networks},'' {\em IEEE Trans. Mob. Comput.}, vol.~24, no.~4, pp.~3441--3455, 2025.

\bibitem{Wang2024}
D.~Wang, X.~Jin, Z.~Feng, and Q.~Deng, ``{B-UFS: Uniform Resource Metric-Based Periodic Flow Scheduling in 5G-TSN Integrated Network},'' in {\em IEEE SIES}, pp.~140--147, 2024.

\bibitem{Ginthor2020}
D.~Ginthör, R.~Guillaume, J.~von Hoyningen-Huene, M.~Schüngel, and H.~D. Schotten, ``{End-to-end Optimized Joint Scheduling of Converged Wireless and Wired Time-Sensitive Networks},'' in {\em IEEE ETFA}, vol.~1, pp.~222--229, 2020.

\bibitem{Chen2022}
Y.~Chen, X.~Yao, Z.~Gan, Z.~You, P.~Wu, and W.~Wang, ``{Power Service Mapping Scheduling Method Based on Fusion of 5G and Time-Sensitive Network},'' in {\em IEEE IAEAC}, pp.~1309--1314, 2022.

\bibitem{Shih2023}
Y.-Y. Shih, H.-C. Liu, C.-C. Chuang, and A.-C. Pang, ``{Scheduling of Integrated 5G and Time Sensitive Network for Deterministic Communication},'' in {\em IEEE ETFA}, pp.~1--8, 2023.

\bibitem{Fon24}
J.~Fontalvo-Hernández, A.~Zirkler, and T.~Bauschert, ``{Determinism in Industrial Converged Networks: Evaluating Approaches to Jitter Mitigation in 5G and TSN Integration},'' in {\em IFIP Networking}, pp.~744--749, 2024.

\bibitem{Jayabal2025}
R.~J. Jayabal, D.~T.~C. Wong, L.~K. Goh, X.~Zhang, C.~M. Pang, and S.~Sun, ``{Survey, Design and Evaluation of TGT-HC: A Time-Aware Shaper MAC for Wireless TSN},'' {\em IEEE Trans. Mob. Comput.}, vol.~24, no.~6, pp.~5433--5445, 2025.

\bibitem{torra24}
A.~Agustí-Torra, M.~Ferré-Mancebo, and D.~Rincón-Rivera, ``{Emulating Integrated 5G-TSN Scenarios},'' in {\em {2024 15th International Conference on Network of the Future (NoF)}}, pp.~96--100, 2024.

\bibitem{Aij24}
A.~Aijaz and S.~Gufran, ``{Time-Sensitive Networking over 5G: Experimental Evaluation of a Hybrid 5G and TSN System with IEEE 802.1Qbv Traffic},'' in {\em NoF}, pp.~101--105, 2024.

\end{thebibliography}
\vspace{-1cm}
\begin{IEEEbiography}
[{\includegraphics[width=1in,height=1.25in,clip,keepaspectratio]{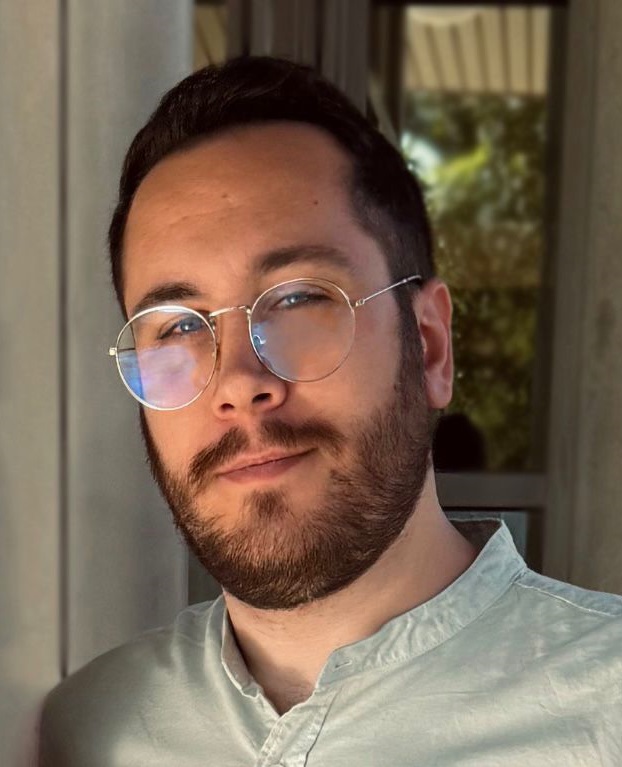}}]{PABLO RODRIGUEZ-MARTIN} obtained his B.Sc. and M.Sc. degrees in Telecommunications Engineering from the University of Granada (UGR), Granada, Spain, in 2020 and 2022, respectively. He is currently pursuing his Ph.D. as a researcher in the WiMuNet Lab Research Group, affiliated to the Department of Signal Theory, Telematics and Communications (TSTC), University of Granada. His research focuses on Time-Sensitive Networking (TSN), 5G/6G networks, Industrial Internet of Things (IIoT), and Artificial Intelligence.
\end{IEEEbiography}
\vspace{-1cm}
\begin{IEEEbiography}[{\includegraphics[width=1in,height=1.25in,clip,keepaspectratio]{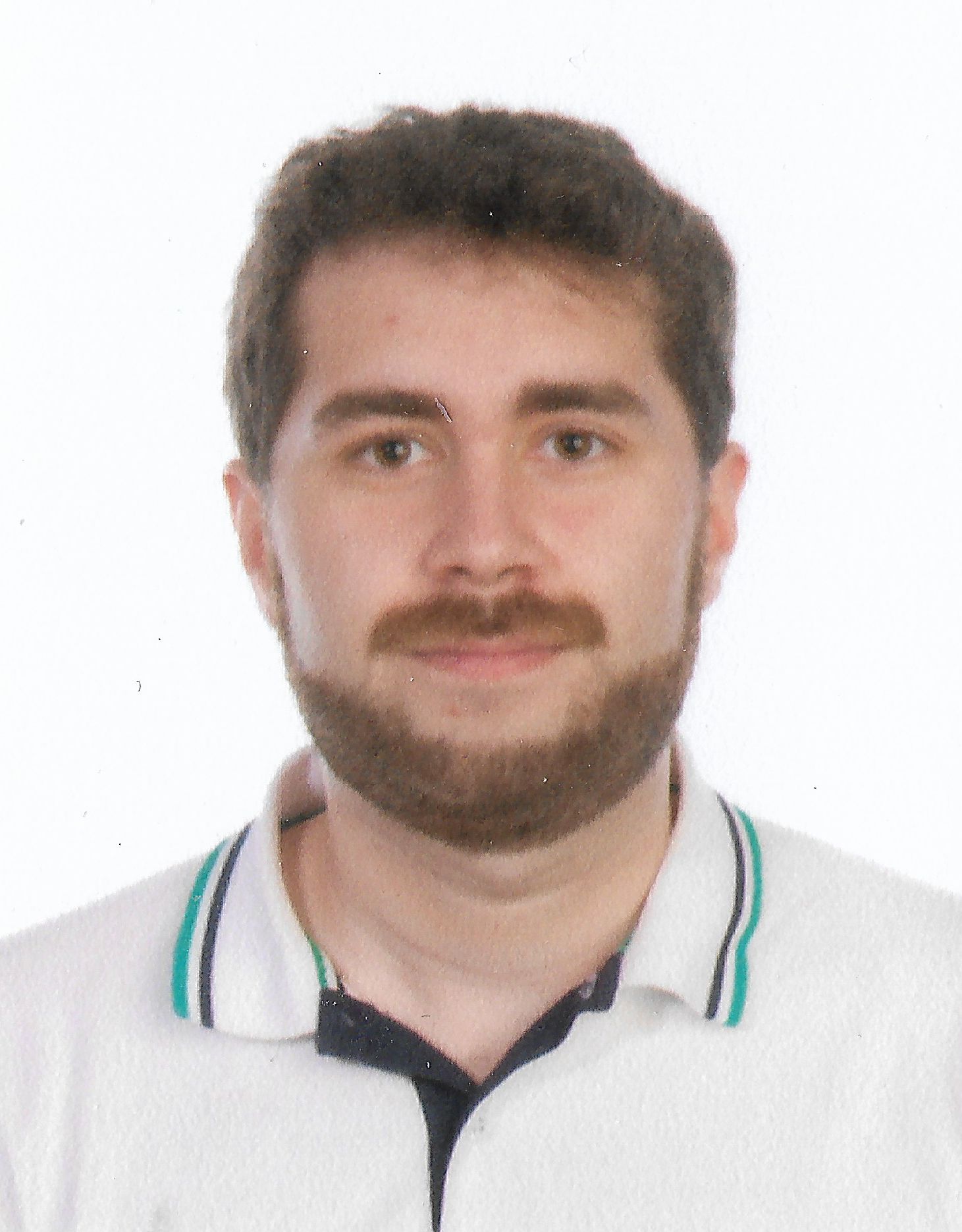}}]{OSCAR ADAMUZ-HINOJOSA} received the
B.Sc., M.Sc., and Ph.D. degrees in telecommunications engineering from the University of Granada (UGR), Granada, Spain, in 2015, 2017, and 2022, respectively. He was granted a Ph.D. fellowship by the Education Spanish Ministry in September 2018. He is currently an Interim Assistant Professor with the Department of Signal Theory, Telematics, and Communication (TSTC), University of Granada. He has also been a Visiting Researcher at NEC Laboratories Europe on several occasions. His research interests include network slicing, 6G Radio Access Networks (RAN), and deterministic networks, focusing on mathematical modeling.
\end{IEEEbiography}
\vspace{-1cm}
\begin{IEEEbiography}[{\includegraphics[width=1in,height=1.25in,clip,keepaspectratio]{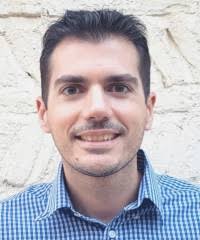}}]{PABLO MUÑOZ} received the M.Sc. and Ph.D. degrees in telecommunication engineering from the University of Málaga (UMA), Málaga, Spain, in 2008 and 2013, respectively. He is currently an Associate Professor with the Department of Signal Theory, Telematics, and Communications (TSTC), University of Granada (UGR), Granada, Spain. He has published more than 50 articles in peer-reviewed journals and conferences. He is the coauthor of four international patents. His research interests include Radio Access Network (RAN) planning and management, the application of Artificial Intelligence (AI) tools in Resource Block (RB) management, and the virtualization of wireless networks.
\end{IEEEbiography}
\vspace{-1cm}
\begin{IEEEbiography}[{\includegraphics[width=1in,height=1.25in,clip,keepaspectratio]{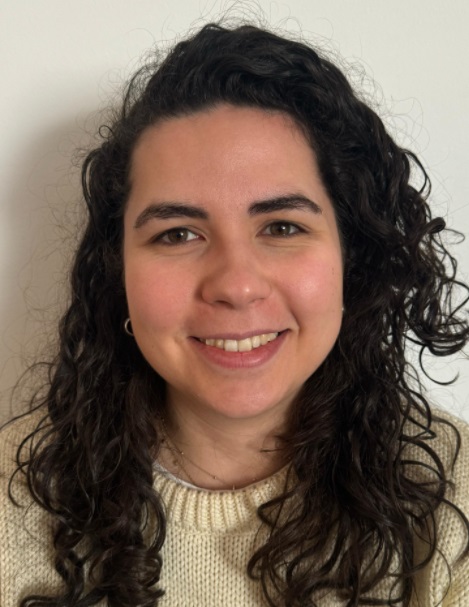}}]{JULIA CALEYA-SANCHEZ} received the B.Sc. and M.Sc. degrees from the University of Granada (UGR), Granada, Spain, in 2021 and 2023, respectively. She was granted a Ph.D. fellowship by the Education Spanish Ministry in September 2022. She is currently pursuing the Ph.D. degree with the WiMuNet Laboratory Research Group, Department of Signal Theory, Telematics and Communications (TSTC), University of Granada. Her research interests include Time-Sensitive Networking (TSN) and Industry 4.0.
\end{IEEEbiography}
\vspace{-1cm}
\begin{IEEEbiography}[{\includegraphics[width=1in,height=1.25in,clip,keepaspectratio]{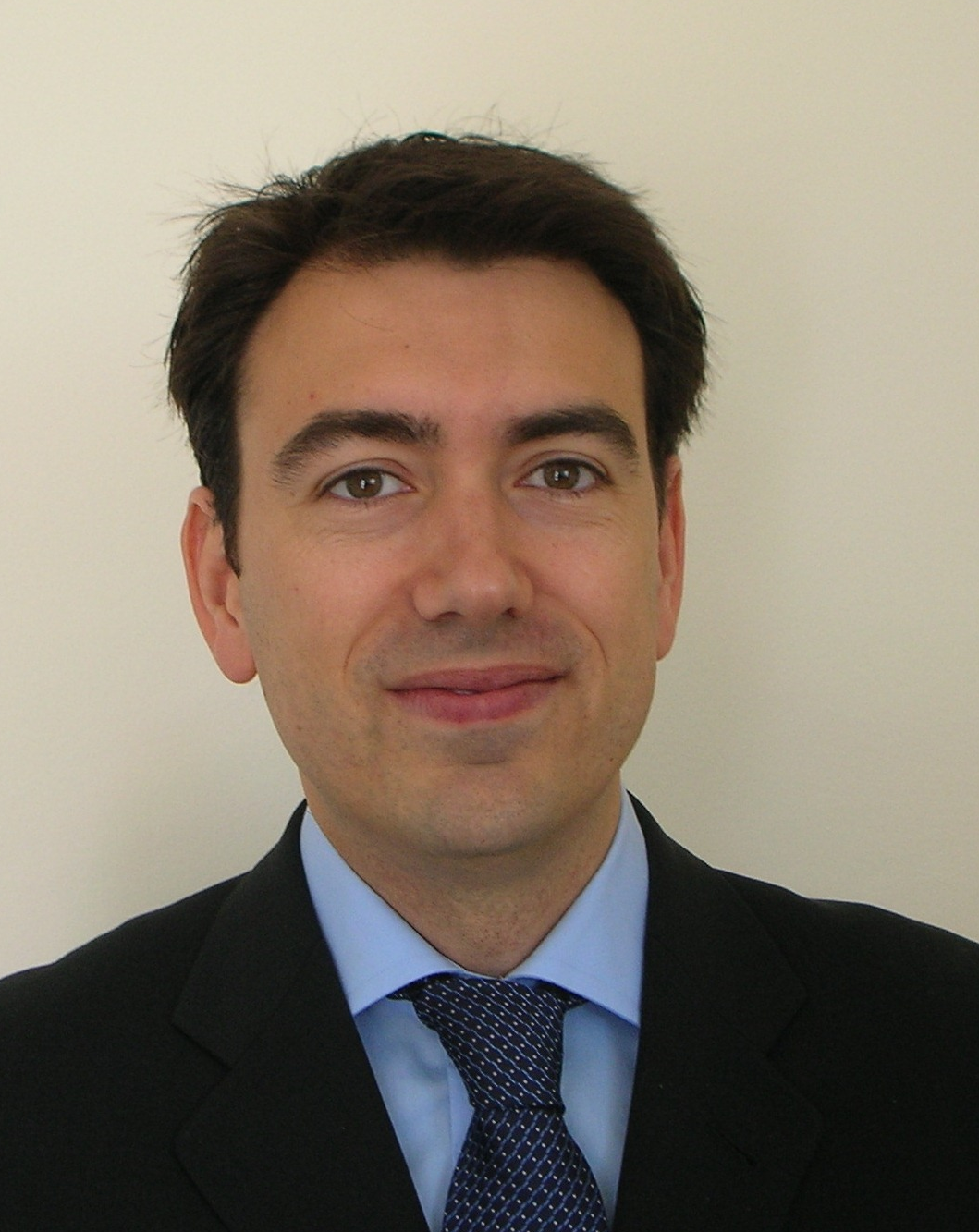}}]{PABLO AMEIGEIRAS} received the M.Sc.E.E. degree from the University of Málaga (UMA), Málaga, Spain, in 1999. He carried out his Master thesis at the Chair of Communication Networks, Aachen University (RWTH), Aachen, Germany. In 2000, he joined Aalborg University (AAU), Aalborg, Denmark, where he carried out his Ph.D. thesis. In 2006, he joined the University of Granada (UGR), Granada, Spain, where he has been leading several projects in the field of 4G and 5G systems. He is currently a Full Professor at the Department of Signal Theory, Telematics and Communications (TSTC). His research interests include 5G, 6G, the Internet of Things (IoT), and deterministic networks.
\end{IEEEbiography}
\end{document}